# The Formal Semantics and Implementation of a Domain-Specific Language for Mixed-Initiative Dialogs

Zachary S. Rowland[a]  and Saverio Perugini[b]

a    Department of Computer Science, University of Dayton, USA
b    Department of Mathematics, Ave Maria University, USA

**Abstract**   Human-computer dialog plays a prominent role in interactions conducted at kiosks (e.g., withdrawing money from an ATM or filling your car with gas), on smartphones (e.g., installing and configuring apps), and on the web (e.g., booking a flight). Some human-computer dialogs involve an exchange of system-initiated and user-initiated actions. These dialogs are called *mixed-initiative dialogs* and sometimes also involve the pursuit of multiple interleaved sub-dialogs, which are woven together in a manner akin to coroutines. However, existing dialog-authoring languages have difficulty expressing these dialogs concisely. In this work, we improve the expressiveness of a dialog-authoring language we call *dialog specification language* (DSL), which is based on the programming concepts of functional application, partial function application, currying, and partial evaluation, by augmenting it with additional abstractions to support concise specification of task-based, mixed-initiative dialogs that resemble concurrently executing coroutines. We also formalize the semantics of DSL—the process of simplifying and staging such dialogs specified in the language. We demonstrate that dialog specifications are compressed by to a higher degree when written in DSL using the new abstractions. We also operationalize the formal semantics of DSL in a Haskell functional programming implementation. The Haskell implementation of the simplification/staging rules provides a proof of concept that the formal semantics are sufficient to implement a dialog system specified with the language. We evaluate DSL from practical (i.e., case study), conceptual (i.e., comparisons to similar systems such as VoiceXML and State Chart XML), and theoretical perspectives. The practical applicability of the new language abstractions introduced in this work is demonstrated in a case study by using it to model portions of an online food ordering system that can be concurrently staged. Our results indicate that DSL enables concise representation of dialogs composed of multiple concurrent sub-dialogs and improves the compression of dialog expressions reported in prior research. We anticipate that the extension of our language and the formalization of the semantics can facilitate concise specification and smooth implementation of task-based, mixed-initiative, human-computer dialog systems across various domains such as ATMs and interactive, voice-response systems.

## The Art, Science, and Engineering of Programming

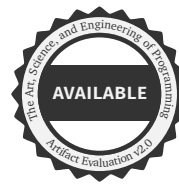
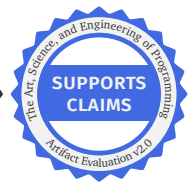

# The Formal Semantics and Implementation of a DSL for Mixed-Initiative Dialogs

> As Achilles suggested, perhaps the desired information lies "closer to the surface" in one representation than in another.
> — Douglas R. Hofstadter, *Gödel, Escher, Bach: An Eternal Golden Braid* [19, p. 427]

## 1 Introduction

The use of human-computer dialog[1] plays a prominent role in user interfaces for common tasks such as withdrawing money from an ATM, installing and configuring smart phone apps, or booking a flight [9, 28]. Improvements in natural language processing and artificial intelligence have led to the development of conversational interfaces and personal assistants (e.g., Apple's Siri and Amazon's Alexa), which utilize natural language.

The number of ways to complete a dialog, all of which must be supported by an implementation, is a general measure of the flexibility afforded to the user. Consider a dialog where the user has four opportunities to steer the dialog in one of three ways. Implementing this dialog requires modeling and managing $81$ $(= 3^4)$ paths to dialog completion. That number increases exponentially as the user is afforded more opportunities for taking the dialog initiative [10]. A *mixed-initiative* dialog is a dialog in which both the user and system exchange initiative to steer the flow of the dialog toward (task) completion [2, 30, 41]. In addition, the inherently ambiguous nature of natural language adds complexity to dialog management [23, 42] and often requires the system to insert clarification questions and, when necessary, to backtrack and redo parts of the dialog as responses are progressively provided and discerned. Thus, writing a program that facilitates a rich form of human-computer dialog is time-consuming and error-prone [28].

Dialog-authoring languages and management systems facilitate the construction of a dialog system [9, 28]. The programmer authors a dialog specification, which the dialog manager then caries out [23]. Our group has developed a dialog-authoring language we call *Dialog Specification Language* (DSL hereafter) for task-based, mixed-initiative dialog specification and implementation [31, 33] whose semantics are inspired by the programming language concepts of functional application (i.e., interpretation), partial function application, currying [32, Sections 4.3, 8.2, 8.3, respectively], and partial evaluation [21].

In prior work we have demonstrated that these programming language concepts correspond to paths (en route) to dialog completion in a spectrum from completely fixed to completely flexible (i.e., mixed-initiative) dialog paths [31, 33]; these concepts enable a *concise* specification of task-based, mixed-initiative dialogs [31]; and this alternate perspective on dialog specification leads to dialog implementation with attractive implications on dialog management [7, 33]. Thinking of paths through a dialog as the (partial) application of (curried) function or the partial evaluation of a function constitutes the central *artistic* motif of this article/work.

---

[1] *Human-computer dialog* in this setting refers to any sequence of interactions between a user and a system (e.g., a user completing an interactive application for an insurance quote), not necessarily pursued through a verbal modality.

This paper extends that prior work in two distinct ways. First, we extend DSL to address one specific modeling limitation of the language—specifying dialogs involving concurrent tasks in a *concise* way (Sections 2.8 and 3.3–3.4). Secondly, we formalize the semantics of the entire language (Section 3), including the syntactic extensions described here. We operationalize those formal semantics as a proof of concept in a functional programming implementation in Haskell (Section 3.5)—enhancing the *artistic* aspect of this work and the relevance of this article to this journal.

A goal of DSL is to define a dialog specification that captures the permissible paths through a dialog more concisely than possible with other representations (e.g., finite-state machines, VoiceXML, State Chart XML). The expressiveness of DSL informally refers to its ability to capture all the permissible paths through the dialog using the least number of expressions from the language. In this way, DSL can be viewed as losslessly compressing the permissible paths, and the expressiveness of the language can be thought of as the magnitude of the compression possible. DSL does not have an abstraction for concisely specifying dialogs that support the interruption (and later resumption) of a task in favor of another, more immediate, task (i.e., concurrent tasks). Our extension of DSL enables the language to capture more of the dialogs in the spectrum of dialogs described above [31]. The syntactic additions enable concise representation of dialogs that exhibit concurrent-task behavior (Section 2.8) and the formalization of the semantics (Section 3) provides a baseline against which implementations (Sections 3.5) can be verified for correctness.

## 2 Dialog Specification Language

*Dialog Specification Language* (DSL) is a dialog-authoring language for task-based, mixed-initiative dialog specification and implementation [31, 33] whose semantics are inspired by the programming language concepts of functional application (i.e., interpretation), partial function application, currying [32, Sections 4.3, 8.2, 8.3, respectively], and partial evaluation [21]. In this section we focus on the specification part of the language. (Section 3 focuses on the semantics and implementation of the language.) The main motif of the language is that these programming language concepts correspond to paths (en route) to dialog completion in a spectrum from completely fixed to completely flexible (i.e., mixed-initiative) dialog paths [31, 33]. We explore this essential motif of the language with some concrete examples.

The form of a DSL expression is $\frac{\mathcal{M}}{e_1\ e_2\ \cdots\ e_n}$, where $\mathcal{M}$ represents a mnemonic for a programming language concept and $e_1 e_2 \ldots e_n$ are DSL (sub-)expressions. For now, we can think of each sub-expression simply as a prompt for user input we call a *solicitation*, denoted by a string of characters (e.g., credit-card, octane, receipt? in a gas-kiosk dialog). Each concept mnemonic specifies a set of allowable orderings in which the solicitations in the denominator can be pursued.

## 2.1 Completely Fixed Order: Currying

The $C$ mnemonic refers to a *curried function* [32, Section 8.3] and, thus, the expression $\frac{C}{e_1\ e_2\ \cdots\ e_n}$ requires its $n$ solicitations to be pursued in a *strict order* from left to right and *one at a time* akin to the complete evaluation of a curried function. Consider a dialog for purchasing gas at a kiosk:

$$\frac{C}{\text{credit-card\ \ octane\ \ receipt?}} \equiv \{\prec\text{credit-card\ \ octane\ \ receipt?}\succ\}$$

The dialog used at the gas-pump kiosk typically proceeds in this fixed order. The right-hand side of the $\equiv$ sign is a set of sequences or paths through the dialog (in this case, there is only one), each of which we call an *episode*.[2] Neither the $\equiv$ sign nor the set of episodes is part of the DSL expression, but is shown to help the reader discern which set of episodes the DSL expression on the left-hand side represents.

## 2.2 Combinations of Multiple Responses Per Utterance: Interpretation

In some interfaces, users are permitted to respond to more than one solicitation in a single utterance. For instance, consider a simple HTML form with multiple textfields and a single submit button for an online mortgage application on a single webpage. In this case, the user is responding to multiple prompts in a single utterance when clicking the submit button. Such user interaction corresponds to a complete application of a function to its arguments (i.e., the individual responses, each of which is mapped onto an individual solicitation, which is the analog of a parameter). We use the mnemonic $I$ (for interpretation in the programming languages context) to refer to this interaction:

$$\frac{I}{\text{salary\ \ credit-score\ \ age}} \equiv \{\prec\{\text{salary, credit-score, age}\}\succ\}$$

Notice that, unlike the prior episode $\prec$credit-card octane receipt?$\succ$, this episode $\prec\{\text{salary, credit-score, age}\}\succ$ uses set curly braces around the solicitations to indicate that the responses to them must be communicated in the same utterance or, in other words, at the same time and, thus, are unordered.

## 2.3 All Orders and All Combinations: Partial Evaluation

The $PE$ mnemonic stands for *partial evaluation* [21] and the dialog expression $\frac{PE^\star}{e_1\ e_2\ \cdots\ e_n}$ contains $n$ solicitations that can be answered *in any order* and *in any combination* of user responses. Consider the following coffee-ordering dialog:

$$\frac{PE^\star}{\text{size\ \ blend\ \ type-of-milk}} \equiv \left\{\begin{array}{c} \prec\text{size\ blend\ type-of-milk}\succ, \prec\text{size\ type-of-milk\ blend}\succ, \prec\text{blend\ size\ type-of-milk}\succ, \prec\text{blend\ type-of-milk\ size}\succ, \\ \prec\text{size\ blend\ type-of-milk}\succ, \prec\text{size\ type-of-milk\ blend?}\succ, \prec\{\text{size, blend}\}\ \text{type-of-milk}\succ, \\ \prec\text{type-of-milk\ \{size, blend\}}\succ, \prec\{\text{size, type-of-milk}\}\ \text{blend}\succ, \prec\{\text{blend, type-of-milk}\}\ \text{size}\succ, \\ \prec\text{size\ \{blend, type-of-milk\}}\succ, \prec\text{blend\ \{size, type-of-milk\}}\succ, \prec\{\text{size, blend, type-of-milk}\}\succ \end{array}\right\}$$

Structuring the interaction for a dialog represented by the expression on the left-hand side can be thought of as repeatedly partially evaluating a function—hence, the star ($\star$) superscript on the mnemonic—until a fixpoint is reached. During partial evaluation, a subset of the parameters of the function are bound to arguments, and the original function is transformed[3] into a function that accepts arguments for the remaining parameters. The total number of ways a function of three parameters can be repeatedly partially evaluated corresponds to the number of episodes specified by this DSL expression (i.e., thirteen). As we see in this example, one DSL expression captures a set of (thirteen) episodes, which we call an *enumerated* (*dialog*) *specification*. The use of partial evaluation for specifying and implementing human-computer dialogs was introduced by Ramakrishnan, Capra, and Pérez-Quiñones [37] who posed it as a method to support *unsolicited reporting* [2], a form of mixed-initiative interaction.

---

[2] See Appendix A for a list of terms and their definitions used in this article.

### 2.4 Sub-Dialogs

The $C$ and $PE^\star$ type dialogs represent two ends of a spectrum of task-based, mixed-initiative dialogs [31, Table I]: those with one fixed order of solicitations ($C$) and those supporting all possible orders and all possible combinations of solicitations ($PE^\star$), respectively. One way to capture more of the space between these two ends is to introduce sub-dialogs. DSL expressions can be nested (within the denominator) to specify more complex dialogs (e.g., those involving sub-dialogs). As a first example consider the following DSL expression:

$$\cfrac{C}{\text{size} \quad \cfrac{I}{\text{blend} \quad \text{type-of-milk}}} \equiv \{\prec\text{size} \ \{\text{blend}, \ \text{type-of-milk}\}\succ\}$$

This expression specifies a dialog with one episode containing two ordered utterances: the first containing a single response to the solicitation for size[4] and the second containing two responses—one for blend and one for type-of-milk—communicated in one stroke in a single utterance. Thus, we have relaxed the above restriction on the (sub-)expressions in the denominator of a DSL expression from simple, atomic solicitations for user input (e.g., $\cfrac{C}{\text{credit-card} \quad \text{octane} \quad \text{receipt?}}$) to any combination of atomic solicitations and DSL (sub-)expressions (e.g., $\cfrac{C}{\text{size} \quad \cfrac{I}{\text{blend} \quad \text{type-of-milk}}}$) As a second example consider a dialog to book a flight:

$$\cfrac{C}{\text{departure-time} \quad \cfrac{PE^\star}{\text{from} \quad \text{to}} \quad \text{seat}} \equiv \left\{ \begin{array}{l} \prec\text{departure-time} \quad \text{from} \quad \text{to} \quad \text{seat}\succ, \\ \prec\text{departure-time} \quad \text{to} \quad \text{from} \quad \text{seat}\succ, \\ \prec\text{departure-time} \ \{\text{from, to}\} \ \text{seat}\succ \end{array} \right\}$$

Again, here one DSL expression captures an enumerated specification consisting of (three) episodes.

### 2.5 All Possible Orderings Only: Stepwise Partial Evaluation

At this point it is appropriate to introduce stepwise partial evaluation because of its relevance to staging dialogs involving multiple sub-dialogs. Stepwise partial evaluation

---

[3] Partial evaluation is a source-to-source program transformation [21].
[4] If an utterance contains a single response, the curly braces are omitted.

# The Formal Semantics and Implementation of a DSL for Mixed-Initiative Dialogs

**Table 1** Programming Language Concepts Used in DSL

| Programming Language Concept | Mnemonic | Star* | Prime' | Notes |
|---|---|---|---|---|
| interpretation | $I$ | – | – | – |
| currying | $C$ | – | – | – |
| stepwise partial evaluation | $SPE$ | $SPE^\star$ | $SPE'$ | – |
| partial function application one | $PFA_1$ | $PFA_1^\star$ | $PFA_1'$ | $PFA_1' \equiv C$ |
| partial function application $n$ | $PFA_n$ | $PFA_n^\star$ | $PFA_n'$ | $PFA_n^\star \equiv PFA_n'$ |
| partial evaluation | $PE$ | $PE^\star$ | $PE'$ | $PE^\star \equiv PE'$ |
| pause and resume (weaving) | $W$ | – | – | – |

restricts partial evaluation to one parameter at a time. Thus, while $\frac{PE^\star}{\text{size blend type-of-milk}}$ corresponds to all thirteen ways a function of three parameters can be repeatedly partial evaluated and, thus, captures all possible orderings and groups of the three solicitations, $\frac{SPE'}{\text{size blend type-of-milk}}$ captures all possible ordering of individual solicitations only (i.e., all six permutations of the solicitations). While $C$ only supports fixed orders of single-response utterances, $SPE'$ only supports flexible orders of single-response utterances.

## 2.6 Other Concepts from Programming Languages in DSL

There are fifteen concept mnemonics in DSL (Table 1). There are seven base mnemonics (in the **Mnemonic** column), four of which have star and prime variants. Three are extraneous because the semantics of each is equivalent to the semantics of another (see **Notes** column), which means that twelve are used in practice.

We have described the other concept mnemonics in other articles and refer readers to [33] for the details, especially since those mnemonics not significantly relevant to the primary focus of this paper, which is the formal semantics of the language, including the $W$ mnemonic and ↯ operator. However, it is helpful at this point in our discussion to distinguish between the star and prime mnemonic variants. The star ($\star$) superscript on a mnemonic *permits*, but does not require, repeated applications of the mnemonic operator; the prime ($'$) superscript *requires* repeated applications until a fixpoint is reached. The star-variant mnemonics permit the user to complete the dialog at any time by responding to the remaining solicitations in a single utterance, which is not permitted by the prime-variant mnemonics. For instance, the episode {≺size {blend, type-of-milk}≻ is captured by DSL expression $\frac{SPE^\star}{\text{size blend type-of-milk}}$, but not by the expression $\frac{SPE'}{\text{size blend type-of-milk}}$ [33].

The $SPE'$ mnemonic has particular relevance to modeling dialogs that involve multiple sub-dialogs. Recall that $SPE'$ permits all possible completion orders of solicitations one at a time (i.e., each utterance is restricted to one response). Thus, when $SPE'$ is used in the numerator of a DSL expression whose denominator contains multiple sub-dialogs, the $SPE'$ mnemonic affords the user flexibility in the order in which those sub-dialogs are pursued, while permitting only one sub-dialog to be pursed at a time. In other words, $SPE'$ can be used to specify a dialog with more than two DSL (sub-)expressions

7:6

in the denominator, any of which can be a sub-dialog, that can be completed in any order. Note that

$$\cfrac{C}{\cfrac{PE^*}{\text{size blend}} \; \cfrac{PE^*}{\text{eggs toast}} \; \cfrac{PE^*}{\text{credit-card receipt?}}} \neq \cfrac{SPE'}{\cfrac{PE^*}{\text{size blend}} \; \cfrac{PE^*}{\text{eggs toast}} \; \cfrac{PE^*}{\text{credit-card receipt?}}},$$

but

$$\cfrac{C}{\cfrac{PE^*}{\text{size blend}} \; \cfrac{PE^*}{\text{eggs toast}} \; \cfrac{PE^*}{\text{credit-card receipt?}}} \subset \cfrac{SPE'}{\cfrac{PE^*}{\text{size blend}} \; \cfrac{PE^*}{\text{eggs toast}} \; \cfrac{PE^*}{\text{credit-card receipt?}}};$$

the episode ≺{eggs,toast} receipt? credit-card size blend≻ is supported by the latter and not the former, where there is a fixed-order on the sub-dialogs [33].

### 2.7 Expressiveness or Compression

A goal of DSL is to define a dialog specification that captures the permissible paths through a dialog (i.e., the required set of episodes in an enumerated specification) more concisely than possible with other representations (e.g., finite-state machines, VoiceXML; see Figures 8 and 9). In this way, DSL can be viewed as losslessly compressing the set of episodes, and the expressiveness of it can be thought of as the magnitude of the compression possible. The expressiveness of DSL informally refers to its ability to capture all the permissible paths through the dialog using the least number of expressions from the language. Consider the enumerated specification on the right-hand side below:

$$\cfrac{C}{\text{size} \; \cfrac{I}{\text{blend type-of-milk}}} \cup \cfrac{C}{\text{blend} \; \cfrac{I}{\text{size type-of-milk}}} \cup \cfrac{C}{\text{type-of-milk} \; \cfrac{I}{\text{size blend}}} \equiv \left\{ \begin{array}{l} \prec\text{size \{blend, type-of-milk\}}\succ, \\ \prec\text{blend \{size, type-of-milk\}}\succ, \\ \prec\text{type-of-milk \{size, blend\}}\succ \end{array} \right\}.$$

This specification cannot be represented with a single DSL expression. We use the set union of multiple DSL expressions to represent the union of the episode sets that each specifies individually. Thus, the union of DSL expressions on the left-hand side above captures the (three) requisite episodes. Unions of DSL expressions are used to specify dialogs that cannot be specified in DSL (i.e., with a single DSL expression).

One way of effecting more compression is to augment DSL with additional concepts from programming languages that have analogs in dialog interactions between humans and computers, such as the interruption and resumption of dialog coroutines.

### 2.8 Interruption and Resumption of Sub-Dialogs: Coroutines

Often mixed-initiative, human-computer dialogs involve the interruption (and later resumption) of a task in favor of another, more immediate, task (i.e., concurrent tasks). (Dialogs that support concurrent tasks have applications including communications between humans and autonomous devices [24].) The notion of interruption suggests that dialogs can be thought of as coroutines that pass control to each other to accept user input.[5] (The dialogs discussed in this section were not supported by our prior work [7, 31, 33].)

We want DSL to be able to concisely capture specifications of dialogs that involve a weaving of distinct conversational coroutines, which exist in the spectrum of dialogs identified above. Alternatively, these dialogs can be viewed as having some restrictions on the order of solicitations, without requiring them to be answered strictly sequentially. For example, consider the following enumerated dialog specification:

$$\left\{\begin{array}{l}\prec\text{call-attendant-for-help name credit-card octane receipt?}\succ,\\ \prec\text{credit-card call-attendant-for-help name octane receipt?}\succ,\\ \prec\text{credit-card octane call-attendant-for-help name receipt?}\succ,\\ \prec\text{credit-card octane receipt? call-attendant-for-help name}\succ\end{array}\right\}$$

This dialog requires answering the solicitations ≺credit-card octane receipt?≻, in that order, and ≺call-attendant-for-help name≻, in that order. However, at any time, the coroutine (i.e., ≺credit-card octane receipt?≻) can be interrupted and the ≺call-attendant-for-help name≻ path or coroutine (i.e., episode) can be pursued instead. Equipped only with the concepts of programming languages discussed until now, DSL is unable to model this dialog without a union of four *c* expressions.

A sub-dialog can be thought of as a child coroutine, and the mnemonic can be thought of as the parent coroutine which is responsible for managing the execution of its children. For example, an expression using the *c* mnemonic requires that its children begin and terminate in a strict total order—a sub-dialog must finish completely before the next one can begin. Thus, the parent always passes control to the leftmost non-completed sub-dialog. When a child sub-dialog completes, it passes control back to its parent.

We denote the suspension of a dialog using the symbol ↱ after a solicitation. For example, the previous enumerated specification is captured using *w* and ↱ by the following DSL expression:

$$\cfrac{\cfrac{}{\text{credit-card}^{\text{↱}} \ \text{octane}^{\text{↱}} \ \text{receipt?}}\ c \quad \cfrac{}{\text{call-attendant-for-help name}}\ c}{\phantom{x}}\ w$$

The semantics of the ↱ operator are that after the user responds to the credit-card (or octane) solicitation control is passed one level higher than normal (i.e., to the *w* mnemonic instead of the *c* mnemonic). The *w* stands for "weaving" and specifies that the set of user responses needed to complete sub-dialogs can be woven together. When an expression containing a *w* receives control, it can pass control to *any* of its sub-dialogs.

The *w* mnemonic and the ↱ operator permit dialogs to be specified in DSL that could not be expressed without them (e.g., the dialog above), including our prior work [7, 31, 33], which did not include the *w* mnemonic and the ↱ operator that are introduced here. As another example, the dialog represented by the union of DSL expressions on the left-hand side below can now be expressed in DSL (see right-hand side):

---

[5] *Coroutines* are procedures or routines that cooperate with each other by yielding to each other and are an example of non-preemptive multitasking.

$$\frac{\dfrac{C}{\underset{\text{eggs coffee}}{SPE'}\quad\underset{\text{toast cream?}}{SPE'}}\cup\dfrac{SPE'}{\underset{\text{eggs toast}}{C}\quad\underset{\text{coffee cream?}}{C}}\equiv\dfrac{W}{\underset{\text{eggs}^{\uparrow}\text{ toast}}{C}\quad\underset{\text{coffee}^{\uparrow}\text{ cream?}}{C}}$$

$$\equiv\left\{\begin{array}{l}\prec\text{eggs toast coffee cream?}\succ,\prec\text{eggs coffee toast cream?}\succ,\prec\text{eggs coffee cream? toast}\succ,\\\prec\text{coffee eggs toast cream?}\succ,\prec\text{coffee eggs cream? toast}\succ,\prec\text{coffee cream? eggs toast}\succ\end{array}\right\}$$

This dialog for ordering breakfast involves two sub-dialogs: one for a meal (i.e., type of eggs and choice of toast) and one for a drink (i.e., type of coffee and option of cream or not). The responses to these solicitations can arrive in any order, with the restrictions that choice of toast must not precede type of eggs and, likewise, the choice of cream or not must not precede type of coffee (i.e., episodes such as ≺cream? eggs coffee toast≻ are not permitted).

We can use multiple arrows to specify the passing of control further up the parent-child hierarchy. For instance, consider the following DSL expression:

$$\dfrac{\dfrac{\dfrac{C}{\text{credit-card}^{\uparrow\uparrow}\text{ octane}^{\uparrow}\text{ receipt?}}\quad\text{call-attendant-for-help}}{W}\quad\text{name}}{W}$$

$$\equiv\left\{\begin{array}{l}\prec\text{credit-card octane receipt? call-attendant-for-help name}\succ,\prec\text{credit-card octane receipt? name call-attendant-for-help}\succ,\\\prec\text{credit-card octane call-attendant-for-help receipt? name}\succ,\prec\text{credit-card call-attendant-for-help name octane receipt?}\succ,\\\prec\text{credit-card call-attendant-for-help octane receipt? name}\succ,\prec\text{credit-card name octane receipt? call-attendant-for-help}\succ,\\\prec\text{credit-card name octane call-attendant-for-help receipt?}\succ,\prec\text{credit-card name call-attendant-for-help octane receipt?}\succ,\\\prec\text{call-attendant-for-help credit-card name octane receipt?}\succ,\prec\text{call-attendant-for-help name credit-card octane receipt?}\succ,\\\prec\text{call-attendant-for-help credit-card octane receipt? name}\succ,\prec\text{name credit-card octane receipt? call-attendant-for-help}\succ,\\\prec\text{name credit-card octane call-attendant-for-help receipt?}\succ,\prec\text{name call-attendant-for-help credit-card octane receipt?}\succ,\\\prec\text{name credit-card call-attendant-for-help octane receipt?}\succ\end{array}\right\}$$

Here the solicitation name can be responded to between solicitations credit-card and octane because the $\uparrow\uparrow$ operator returns to the outer $w$, but not between solicitations octane and receipt? because the $\uparrow$ operator returns to the inner $w$.

## 2.9 Formal DSL Syntax and Restrictions on DSL Expressions

Now that we have discussed DSL in an informal, intuitive way, we present the formal syntax of the language (Figure 1),[6] including a meta-notation for specifying dialog patterns (bottom) used in Section 3. While DSL is a context-sensitive language (due to the restrictions on some of the numerator/denominator combinations identified below), it is defined using a context-free grammar. Expressions include the empty dialog (i.e., the dialog with no episodes), denoted ∼; the atomic dialog (i.e., the dialog with only a single solicitation (e.g., size); two types of compound dialogs: those without sub-dialogs and those with sub-dialogs; and unions of compound dialogs (e.g., $\dfrac{C}{\text{size blend}}\cup\dfrac{C}{\text{blend size}}$).

---

[6] Note the overload use of the ⋆ symbol in this grammar. The ⋆ symbol is literal when appearing as the superscript of a concept mnemonic (e.g., $PE^{\star}$). The ⋆ symbol is has its special meaning in EBNF (i.e., zero or more of the previous symbol) when appearing as the superscript of an arrow (e.g., $\uparrow^{\star}$)

**EBNF Grammar of DSL Expressions**

| | | | |
|---|---|---|---|
| <expression> | ::= | ~ | (empty dialog) |
| <expression> | ::= | <solicitation> | (atomic dialog) |
| <expression> | ::= | $\frac{O^{\uparrow\star}}{<solicitations>}$ | (dialog without sub-dialog(s)) |
| <expression> | ::= | $\frac{<M>^{\uparrow\star}}{<expressions>}$ | (dialog with sub-dialog(s)) |
| <expression> | ::= | <expression> ∪ <expression> | (a union of dialogs) |
| <solicitations> | ::= | <solicitation>$^{\uparrow\star}$ | (sub-expression(s) |
| <solicitations> | ::= | <solicitation>$^{\uparrow\star}$ <solicitations> | representing atomic dialog(s)) |
| <expressions> | ::= | <expression> | (sub-expression(s) |
| <expressions> | ::= | <expression> <expressions> | representing (sub)-dialog(s)) |
| <M> | ::= | $C \mid SPE' \mid W \mid PFA_1 \mid SPE$ | |
| <O> | ::= | $I \mid PE \mid PE^\star \mid PE' \mid SPE^\star$ | |
| <O> | ::= | $PFA_1^\star \mid PFA_1' \mid PFA_n \mid PFA_n^\star \mid PFA_n'$ | |

**Meta-Notation for DSL Dialog Patterns**

| | |
|---|---|
| $d$ | Denotes an arbitrary DSL expression. |
| $x$ and $y$ | Denote arbitrary solicitations. |
| $d^\star$ and $x^\star, y^\star$ | Denotes an ordered list of zero or more dialogs or solicitations, respectively. |
| | (e.g., $\frac{C}{x\ d^\star}$ Denotes any DSL expression using the $C$ mnemonic that starts with a solicitation $x$ followed by zero or more expressions representing sub-dialogs) |
| $\{x^\star\}$ | Denotes a single utterance containing responses to multiple solicitations. |

**Figure 1** (top) EBNF grammar of DSL expressions and (bottom) meta-notation for specifying dialog patterns.

There are some context-sensitive, syntactic restrictions on the form of DSL expressions because they have no corresponding semantics.[7] First, sub-dialogs cannot appear under all concept mnemonics. Consider dialogs containing two or more sub-expressions (in the denominator), where at least one of the sub-expressions represents a sub-dialog (e.g., $\frac{SPE'}{\text{size}\ \frac{C}{\text{blend type-of-milk}}}$ and $\frac{SPE'}{\frac{C}{\text{size blend}}\ \frac{PE^\star}{\text{eggs toast}}}$). In such dialogs, none of the $I$, $PFA_n$, $PFA_n^\star$, $PE$, and $PE^\star$ and $SPE$, $SPE^\star$, $PFA_1$, $PFA_1^\star$, $PFA_n'$, or $PE'$ mnemonics may appear in the numerator because those mnemonics require (as in $I$) or support (as in the rest) multiple responses per utterance and it is impossible to complete a sub-dialog in a single utterance. On the other hand, the $PFA_1$ and $SPE$ mnemonics are sufficient for two categories of dialogs containing sub-dialogs: those with no more than two sub-expressions (in the denominator), where one of the sub-expressions represents a sub-dialog (e.g., $\frac{PFA_1}{\text{size}\ \frac{PE^\star}{\text{blend type-of-milk}}}$, $\dfrac{PFA_1}{\dfrac{PE^\star}{\text{size blend}} \quad \dfrac{PE^\star}{\text{eggs toast}}}$, $\dfrac{SPE}{\text{size} \dfrac{PE^\star}{\text{blend type-of-milk}}}$, and $\dfrac{SPE}{\dfrac{PE^\star}{\text{size blend}} \quad \dfrac{PE^\star}{\text{eggs toast}}}$), and those with more than two sub-expressions in the denominator, where only the first sub-expression represents a sub-dialog (e.g., $\dfrac{PFA_1}{\dfrac{PE^\star}{\text{size blend}} \text{ eggs toast credit-card receipt?}}$ and $\dfrac{SPE}{\dfrac{PE^\star}{\text{size blend}} \text{ eggs toast credit-card receipt?}}$). When used as the numerator in an expression whose denominator contains more than two sub-expressions, one of which represents a sub-dialog not in the first position, $PFA_1$ and $SPE$ require multiple responses in the second and final utterance [33]. The $C$, $SPE'$, and $W$ mnemonics are the only mnemonics that can *always* be used (i.e., irrespective of context) in the numerator of an expression containing any arbitrary number of (sub)-expressions representing sub-dialogs in the denominator.

---

[7] In the following discussion we use solicitations labeled $x$ and $y$ for ease of exposition, especially since the purpose of this section is to demonstrate modeling capabilities vis-à-vis the set of episodes supported and not to demonstrate concrete, real-world applications.

Second, only DSL expressions using the $W$ mnemonic are allowed to receive control from ↓.[8] This is to preserve the semantics of the other concept mnemonics. Arrows cannot be placed beneath mnemonics that allow responses to multiple solicitations to be supplied in a single utterance (e.g., $I$ or $PE^\star$). For example, the expression $\dfrac{PE^\star}{\text{credit-card}^\downarrow \text{ octane}^{\downarrow\downarrow} \text{ receipt? rewards}}$ is syntactically invalid because it is unclear which arrows (if any) should apply if a single user utterance responds to the {credit-card, octane, receipt?} set of solicitations.

Despite these restrictions, the ↓ operator has a well-defined behavior in a few non-obvious settings. For instance, consider the following DSL expression (on the left-hand side):

$$\dfrac{W}{\dfrac{SPE'}{\text{size}^\downarrow \text{ blend}} \text{ type-of-milk}} \equiv \left\{ \begin{array}{l} \prec \text{size type-of-milk blend} \succ, \prec \text{size blend type-of-milk} \succ, \\ \prec \text{blend size type-of-milk} \succ, \prec \text{type-of-milk size blend} \succ, \prec \text{type-of-milk blend size} \succ \end{array} \right\}$$

The ↓ operator is applied to the size solicitation if the user responds to it first (see episode ≺size type-of-milk blend≻ on the right-hand side); otherwise it has no effect.

Lastly, an entire DSL expression representing a sub-dialog can be annotated with a ↓ by positioning the ↓ as a superscript of the mnemonic itself (e.g., $\dfrac{I^\downarrow}{\text{salary age}} \equiv \left(\dfrac{I}{\text{salary age}}\right)^\downarrow$). The ↓ operator is applied as soon as the sub-dialog completes. This technique can be used to denote interruption after a set of user responses has been given in a single utterance, as opposed to just one response.

### 2.10 Modeling Case Study: Chipotle

In practice, DSL can be used to model dialogs, and the episodes therein, in the design and implementation of interactive computing systems. We demonstrate here how the interaction involved in using Chipotle's online order form[9] is modeled using DSL. The main order page presents different facets for the user to customize a meal including

---

[8] This does not mean that the ↓ operator cannot appear directly under a concept mnemonic that is not $W$ (e.g., $\dfrac{W}{\dfrac{SPE'}{\text{size}^\downarrow \text{ blend}} \text{ type-of-milk}}$).

[9] https://www.chipotle.com/order/build/burrito-bowl (accessed 2025-01-21).

**The Formal Semantics and Implementation of a DSL for Mixed-Initiative Dialogs**

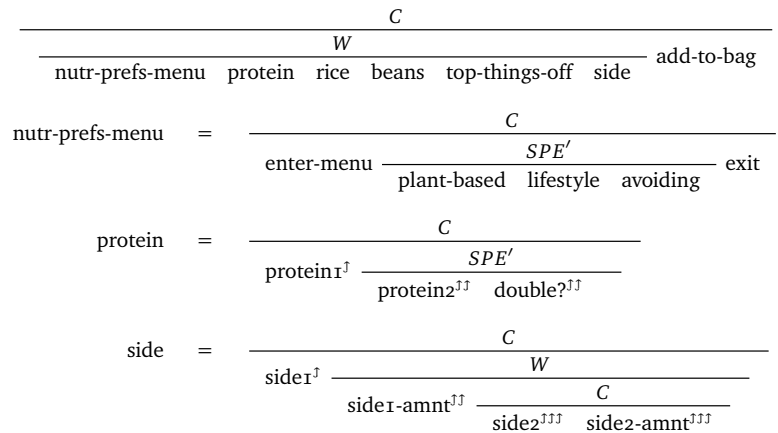

**Figure 2** Dialog specification of the Chipotle online order form in DSL with the $w$ mnemonic.

the type of protein, beans, and rice as well as toppings and side items. Each section has interaction rules governing its completion independent of the others and, thus, corresponds to a different sub-dialog. For instance, the rice section requires the user to make one selection from a set of options (e.g., white rice, brown rice, or no rice). The protein section allows a choice of one or two items from among chicken, steak, carnitas, and other types of protein as well as the option to "double your protein." However, the double option only becomes available once the user has selected at least one type of protein. Also, the toppings section requires a choice of an arbitrary (possibly empty) subset of toppings (e.g., guacamole, salsa, and cheese).

The Chipotle order form provides ample opportunity to make use of the modeling enhancements to DSL introduced here and, thus, the manner in which the user interacts with it provides a testbed in which to utilize the $w$ mnemonic and $\updownarrow$ operator in DSL. With the exception of a few dependencies, the user is not required to complete these sections in the order presented on the page and can jump from section to section. In particular, modeling the sections on the main page for customizing aspects of the order in DSL makes use of the $\updownarrow$ operator since none of these sections require complete control of the interaction. However, the main page also provides access to a nutrition preferences sub-dialog that opens a pop-up window that, when clicked, prevents the user from continuing with other sections until the user makes a selection and exits. Thus, there is no $\updownarrow$ operator in this sub-dialog expression. Figure 2 presents the entire dialog specification in DSL. This case study cannot be modeled in DSL without the $w$ mnemonic and $\updownarrow$ operator in a single DSL expression. The DSL expressions without the $w$ mnemonic and $\updownarrow$ operator, that when unioned, model this interaction is more than is practical to present here.

**Modeling Limitations** DSL lacks certain features required to model nuanced aspects of the Chipotle order form. For instance, the order form enables unimpeded response altering (e.g., the user can de- and re-select options an unlimited number of times and in any order) and DSL does not provide a mechanism for specifying when and how response altering is permitted. In addition, modeling aspects of the order form require finite-state descriptive power. For example, in the "side" section, the form permits the user to select the quantity of an item using "plus" and "minus" buttons. This selection of quantity must be represented by an atomic expression (e.g., "side1-amnt") in DSL. Without direct support for these interactions in DSL, specialized code can be added to the implementation to support these features, which is the approach we have taken in our prior work [6, 33].

## 3 Formal Semantics and Implementation of Dialog Specification Language

In the previous section we discussed the dialog-modeling capabilities of DSL. In this section, we discuss the implementation of the language. We begin by introducing the notion of "staging a dialog"—given a DSL expression and a user utterance, return a new DSL expression representing the new state of the dialog after having processed the responses in the utterance with respect to the previous dialog state. In this way, staging a dialog corresponds to a series of transformations of DSL expressions. We want the DSL expression representing the new dialog state to be as simple as possible, which leads us to simplification rules for rewriting DSL expressions. Both staging and simplification are presented as a set of rewrite rules for DSL expressions that formalize the semantics of the language [3]. Finally, as a proof of concept, we operationalize the rewrite rules in a functional programming implementation in Haskell.

### 3.1 Dialog Staging

*Staging* maps a (DSL expression, utterance) pair to a DSL expression.[10] Formally, staging is a partial function $\mathcal{D}_\mathbb{S} \times \mathbb{U}_\mathbb{S} \rightharpoonup \mathcal{D}_\mathbb{S}$, where $\mathcal{D}_\mathbb{S}$ is the set of dialog expressions with respect to a solicitation-set $\mathbb{S}$, which is the complete set of solicitations posed during a dialog. During dialog staging, the state maintained by the system is implicit in the DSL expression describing the allowable ways to continue the dialog. Intuitively, when the user supplies a valid utterance, the staging system uses it with the current DSL expression to compute a new DSL expression that represents the new dialog state [8, 29]. For example, if the current state is represented by the expression $\dfrac{C}{\text{departure-time} \dfrac{PE^\star}{\text{from to}} \text{first-class?}}$ and the user supplies a response to the solicitation for a departure-time, then the new dialog state is represented by $\dfrac{C}{\dfrac{PE^\star}{\text{from to}} \text{first-class?}}$. The system could also fail for a particular (DSL expression, utterance) pair. For instance,

---

[10] Perugini and Buck [33] focuses on dialog modeling and management only in the context of task-based dialogs. Moreover, user responses (i.e., utterances) are assumed to be structured, well-defined, and unambiguous; system responses are ignored or assumed to be simple confirmation messages. The extension and formalization of that work presented here (i.e., the control unit) also relies on those assumptions and, thus, must be combined with a presentation module and an application interface to be used to implement dialog systems (Figure 3—top).

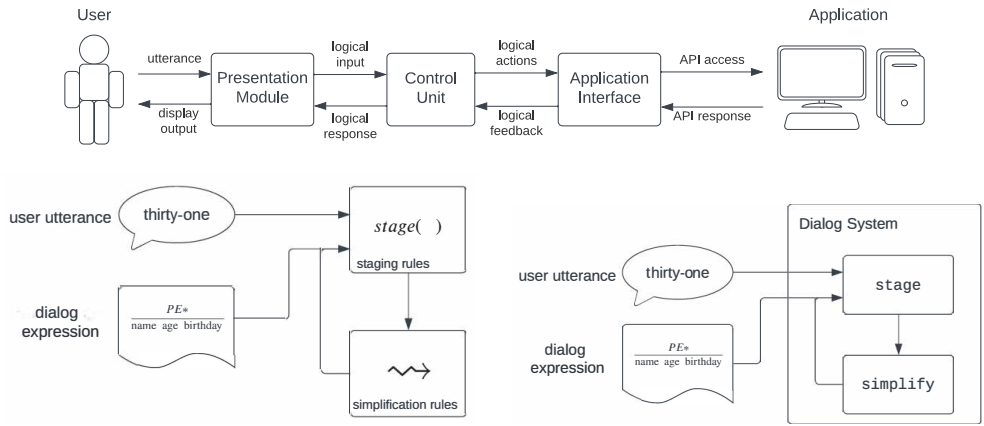

**Figure 3** (top) The Seeheim model for dialog systems [15, 34]. (bottom left) Executing a dialog specified in DSL. (bottom right) General architecture of a dialog system both executing dialogs specified in DSL and built using the implementation of the staging and simplification rules described in Section 3 and Appendix B.

staging will fail if a user responds with a (forthcoming) solicitation for from airport instead of the current solicitation for departure-time.

### 3.2 Rewrite Rules for DSL Expressions

The dialog staging semantics are defined with a set of rewrite rules on DSL expressions [8, 29]. The *simplification rules* are used to build a reduction relation over the set of DSL expressions $\mathcal{D}$ that define a canonical expression for each set of episodes. The *staging rules* define how the dialog state changes when a user supplies an utterance. Since the entire execution state of the dialog system is represented using a DSL expression, the staging rules define a reduction from (DSL expression, utterance) pairs to other DSL expressions.

The simplification and staging rules operate in tandem to accept user utterances and advance the staging process (Figure 3—bottom left). When an utterance arrives, we search the set of staging rules for a rule that can process it with respect to the current DSL expression (i.e., the current state of the dialog). If no such rule is found, then the utterance is rejected and staging fails. Otherwise, the identified staging rule is applied, followed by a reduction to a canonical form using the simplification rules.

To express the rules, a meta-notation is used to specify a dialog pattern (Table 1—bottom). The lower-case symbol $d$ (possibly with a subscript) represents an arbitrary DSL expression, while the symbols $x$ and $y$ denote solicitations. Super-scripting any such symbol with a star ($\star$) indicates an ordered list of zero or more dialogs or solicitations. For example, $\frac{C}{x\ d^\star}$ refers to any DSL expression using the $c$ mnemonic that starts with a solicitation $x$ followed by zero or more expressions representing sub-dialogs. An utterance containing responses to multiple solicitations is always denoted using curly braces and an asterisk (e.g., $\{x^\star\}$). The tilde symbol $\sim$ denotes the empty DSL expression, which represents the dialog with an empty set of episodes.

$$[\text{EMPTY-1}] \quad \frac{\mathcal{M}}{} \quad \leadsto \quad \sim \quad \forall \mathcal{M}$$

$$[\text{EMPTY-2}] \quad \frac{\mathcal{M}}{d_1^\star \ \sim \ d_2^\star} \quad \leadsto \quad \frac{\mathcal{M}}{d_1^\star \ d_2^\star} \quad \forall \mathcal{M}$$

$$[\text{EMPTY-3}] \quad \sim \cup \ d \quad \leadsto \quad d$$

$$[\text{EMPTY-4}] \quad d \cup \sim \quad \leadsto \quad d$$

$$[\text{ATOM-1}] \quad \frac{\mathcal{M}}{d} \quad \leadsto \quad d \quad \mathcal{M} \in \{C, SPE'\}$$

$$[\text{ATOM-2}] \quad \frac{\mathcal{M}}{x} \quad \leadsto \quad x \quad \mathcal{M} \notin \{C, SPE'\}$$

$$[\text{FLATTEN}] \quad \frac{C}{d_1^\star \ \frac{C}{d_2^\star} \ d_3^\star} \quad \leadsto \quad \frac{C}{d_1^\star \ d_2^\star \ d_3^\star}$$

**Figure 4** A sample of the simplification rules for DSL.

### 3.2.1 Simplification Rules

An expression $d_1$ reduces to expression $d_2$ (denoted $d_1 \leadsto^\star d_2$) if both expressions represent the same set of episodes and $d_1$ is at least as complex as $d_2$. Formally, $\leadsto^\star \subset \mathcal{D} \times \mathcal{D}$ is a partial order relation, defined as the reflexive, transitive, and substructural closure of the single-step reduction relation $\leadsto$, which is defined by a set of simplification rules. The *substructural closure* means that any expression can be reduced by reducing one of its sub-expressions.

Figure 4 presents selected simplification rules; the full set is in Appendix B.1. The [EMPTY-1] and [EMPTY-2] rules handle cases involving the empty DSL expression. Any DSL expression with zero sub-expressions (in the denominator) is equivalent to the empty DSL expression. Also, empty DSL expressions can always be removed from sub-expression lists. These rules are particularly useful when applied to sub-expressions after the use of a staging rule. For example, suppose we stage the DSL expression $\frac{C}{\frac{PE^\star}{\text{size blend type-of-milk}} \ \text{rewards-id receipt?}}$ with the utterance {medium, light roast, skim}, which results in $\frac{C}{\frac{PE^\star}{} \ \text{rewards-id receipt?}}$. A $PE^\star$ expression with no sub-expressions (i.e., $\frac{PE^\star}{}$) can be replaced with $\sim$ through the [EMPTY-1] simplification rule, resulting in $\frac{C}{\sim \ \text{rewards-id receipt?}}$. The empty DSL expression can be removed from under the $C$ mnemonic using the [EMPTY-2] rule, resulting in $\frac{C}{\text{rewards-id receipt?}}$.

The [ATOM-1] and [ATOM-2] rules indicate that any expression with a single sub-expression is equivalent to the sub-expression by itself. For instance, the $\frac{PE^\star}{\text{type-of-milk}}$ sub-expression in $\frac{C}{\frac{PE^\star}{\text{type-of-milk}} \ \text{rewards-id receipt?}}$ can be eliminated using [ATOM-2], resulting in $\frac{C}{\text{type-of-milk rewards-id receipt?}}$. These two rules can be thought of as a single rule, but are defined independently to acknowledge that the $C$ and $SPE'$ mnemonics support sub-expressions representing sub-dialogs while the others only support sub-expressions representing solicitations.

Finally, [FLATTEN] means that nested sub-expressions that use $C$ mnemonics can be flattened into an expression using one $C$ mnemonic. For instance, the expressions $\dfrac{C}{\dfrac{C}{\text{size blend}}\text{type-of-milk}}$ and $\dfrac{C}{\text{size }\dfrac{C}{\text{blend type-of-milk}}}$ can both be simplified to $\dfrac{C}{\text{size blend type-of-milk}}$. There does not exist a similar rule for flattening expressions using $SPE'$ because a DSL expression using $SPE'$ requires any nested sub-dialog to be completed entirely before any of the other sub-dialog or solicitation (represented by its sub-expressions) can be pursued. For example, $\dfrac{SPE'}{\text{rewards-id size blend receipt?}}$ specifies the episode ⊰size rewards-id receipt? blend⊱ while $\dfrac{SPE'}{\text{rewards-id }\dfrac{SPE'}{\text{size blend}}\text{ receipt?}}$ does not.

The *canonical form* of a DSL expression $d \in \mathscr{D}_\mathbb{S}$ is the unique expression $c \in \mathscr{D}_\mathbb{S}$ such that $(d \rightsquigarrow^\star c) \wedge (c \rightsquigarrow^\star c') \implies c = c'\ \forall c' \in \mathscr{D}_\mathbb{S}$. For the notion of canonical form to be coherent, the reduction relation must be proven to be antisymmetric and confluent. The relation is *confluent* if whenever $(d \rightsquigarrow d_1) \wedge (d \rightsquigarrow d_2)$ exist, there exists an expression $c$ such that $(d_1 \rightsquigarrow^\star c) \wedge (d_2 \rightsquigarrow^\star c)$.

Constructing a complete proof of these properties is outside the scope of this article and is best implemented using a mechanized proof system such as Coq.[11] However, we are reasonably confident that these propositions are theorems due to the small number of simplification rules and their predictable behavior. By design, none of the rules increase the complexity of an expression, which supports the hypothesis of antisymmetry. Confluence is more challenging to support intuitively. Nonetheless, there are only a few situations where multiple simplification rules could be applied to a single expression—most of them involve empty DSL expressions. In addition, the simplification rules have parallels to other semantic systems with a notion of "simplest form." Sub-structural reduction in particular is a property of many confluent expression systems such as those from arithmetic and pure functional programming (i.e., without side effect).

### 3.2.2 Staging Rules

Each staging rule takes the form $stage(d_1, x) = d_2$, where $x$ is a user utterance and $d_1$ and $d_2$ are DSL expressions. The syntax $stage(d, x)$ stands for the expression that results from responding to given the DSL expression $d$ with $x$. The utterance $x$ can be either a response to one or more solicitations. In the latter case, the solicitations responded to are written as a set. Collectively, the staging rules define a partial function that is deterministic—each (DSL expression, utterance) pair reduces in either one or zero ways. Moreover, the staging function is recursively defined in the case of DSL expressions containing sub-expressions (representing sub-dialogs), or in cases where a union of more than one DSL expression is required to capture the requisite episodes. If staging fails for a sub-expression representing a sub-dialog, it fails for the entire expression/dialog (i.e., failure of sub-expression staging is propagated upward).

Figure 5 presents selected staging rules; the full set is in Appendix B.2. The [ATOM] rule defines staging for an *atomic dialog*—a dialog that processes one solicitation and

$$
\begin{aligned}
&[\text{ATOM}] & stage(x, x) &= \sim \\
&[\text{UNION}] & stage(d_1 \cup d_2, x) &= stage(d_1, x) \cup stage(d_2, x) & \text{if both exist} \\
&[\text{C}] & stage\left(\frac{C}{d_1\ d_2^\star}, x\right) &= \frac{C}{stage(d_1, x)\ d_2^\star} & \text{if } stage(d_1, x) \text{ exists} \\
&[\text{SPE'}] & stage\left(\frac{SPE'}{d_1^\star\ d_2\ d_3^\star}, x\right) &= \frac{C}{stage(d_2, x)\ \frac{SPE'}{d_1^\star\ d_3^\star}} & \text{if } stage(d_2, x) \text{ exists} \\
&[\text{PE}^\star] & stage\left(\frac{PE^\star}{x^\star}, \{y^\star\}\right) &= \frac{PE^\star}{x^\star - y^\star} & \text{if } y^\star \subseteq x^\star
\end{aligned}
$$

---

[11] https://coq.inria.fr (accessed 2025-01-21).

■ **Figure 5** A sample of the staging rules for DSL.

then terminates. If the user response answers the atomic dialog's solicitation, then staging succeeds and reduces to the empty DSL expression. Otherwise, there is no staging reduction defined and the input is rejected. Dialogs that require a union of more than one DSL expression (to capture the requisite episodes) are staged with with three recursive rules. Both the left- and right-hand DSL expressions of a union are staged concurrently until staging fails for one of the expressions, which causes the union to reduce to the other expression. The staging rules [C] and [SPE'] are both recursive since sub-expressions of these mnemonics can be arbitrary expressions instead of only solicitations representing atomic dialogs. The primary difference between the two is that the sub-expressions directly under a $c$ mnemonic must be processed from left to right whereas the sub-dialogs under a $SPE'$ mnemonic can be staged in any order. Notice that staging an expression that uses $SPE'$ introduces a $c$ mnemonic to force the partially complete sub-dialog to terminate before continuing. This additional constraint is lifted by the $w$ mnemonic and ↥ operator discussed next.

### 3.3 Formalization of $W$ Mnemonic and ↥ Operator

In the DSL semantics presented above, a single sub-expression representing a sub-dialog only specifies the behavior of a local section of the dialog. However, in the presence of arrow, a sub-expression (representing a sub-dialog) affects the entire dialog. For example, if the DSL expression $\frac{C}{\text{size}^\uparrow\ \text{blend}}$ appears as a sub-expression, then size and blend need not be answered consecutively. Thus, to support the desired semantics of the ↥ operator within the rewrite-rule reduction model, the DSL expression requires additional rewriting. To illustrate, consider the following example.

$$
stage\left(\frac{W}{\frac{C}{\text{rewards-id credit-card octane}^\uparrow\ \text{car-wash?\ receipt?}}\ \frac{C}{\text{call-attendant\ name}}}, 2391375551\right) = \frac{C}{\text{credit-card\ octane}\ \frac{W}{\frac{C}{\text{car-wash?\ receipt?}}\ \frac{C}{\text{call-attendant\ name}}}}
$$

Since the DSL expression uses the $w$ mnemonic, the user can pursue either of the sub-dialogs before completing the other. However, if the user initially responds with 1987, then the user must continue in the left sub-dialog for the second and third responses. The sub-expression $\frac{C}{\text{credit-card octane}^\uparrow\ \text{car-wash?\ receipt?}}$ must be decomposed to accommodate this new requirement.

The Formal Semantics and Implementation of a DSL for Mixed-Initiative Dialogs

Rewinding a DSL expression to a previous state runs contrary to the use of stateless rewrite rules, where there is no provision for saving an expression after each application of a reduction rule. While techniques are known for defining history-preserving, continuation operations in a reduction framework [12], we prefer a stack-based formulation because of the simplicity of a stack vis-à-vis a first-class continuation.

### 3.3.1 Stack-Based Solution

We define the formal semantics of the $w$ mnemonic and $\uparrow$ operator using a stack-based semantics. In particular, we use a stack to remember how far down the sub-dialog hierarchy the stager has traversed. For example, if the dialog $\dfrac{\dfrac{C}{\text{eggs toast}} \quad \dfrac{C}{\text{coffee cream?}}}{W}$ is staged with the utterance scrambled, then the stager must continue with the sub-dialog $\dfrac{C}{\text{toast}}$ before returning back up to $w$. A stack can remember this information by extracting and re-inserting sub-dialogs at different times.

The staging algorithm operates on a compound structure we call a *reduction state*, which consists of a DSL expression representing the current state of the dialog; an input string; and a stack of dialog constructors. A *dialog constructor* is a function of type Dialog → Dialog that performs the operation of inserting its argument underneath a parent dialog (e.g., $\lambda D. \dfrac{C}{\text{credit-card } D \text{ octane}}$).

This stack-based solution also enables us to unify the two-step semantics (Section 3.2) into a single-step semantics of all mnemonics, including $w$ (Section 3.4). This also has the effect of simplifying the orginal two-step semantics. In particular, only the simplification rules (Figure 4) that involve the empty dialog are actually needed (i.e., [EMPTY-1], [EMPTY-2], [EMPTY-3], and [EMPTY-4]) in this unified semantics. These four simplification rules are used (as demonstrated next) to remove ~ from underneath the $c$ mnemonic.

### 3.3.2 Staging Example

Consider the following initial configuration of the stack (left of double line); DSL expression (right of double line); and input string (outside of box):[12]

| $\lambda D.\ \sim$ ‖ $\dfrac{\dfrac{C}{\text{call-attendant name}} \quad \dfrac{C}{\text{credit-card octane}^{\uparrow}\ \text{receipt?}}}{W}$ | 1234 93 call Joseph yes |

The action taken depends on the shape of the current expression. For expressions using the $w$ mnemonic, the expression is divided into a sub-expression and a dialog constructor. Applying the dialog constructor to the sub-expression must result in the current expression. The sub-expression becomes the current expression, and the dialog constructor is pushed onto the stack. This transformation of the state models the action of entering or extracting a sub-dialog. During the execution of an actual staging algorithm, the correct sub-dialog to extract from underneath $w$ might not be known, and can be non-deterministically chosen. For now, we assume that the staging algorithm knows that extracting the appropriate sub-dialog leads to success.

---

[12] This example uses $\lambda$-calculus notation [26, Section 3.6] for a $\lambda$ function that accepts a dialog $D$ as an argument.

$$\lambda D. \sim \left\| \lambda D. \dfrac{\dfrac{W}{C}}{\text{call-attendant name}} D \right\| \dfrac{C}{\text{credit-card octane}^\uparrow \text{ receipt?}} \quad \text{1234 93 call Joseph yes}$$

For expressions using the *c* mnemonic, we extract the leftmost sub-dialog and, thus, unlike *w* mnemonic, there is no ambiguity as to which sub-dialog to extract since *c* mnemonics must stage their sub-dialogs from left to right.

$$\lambda D. \sim \left\| \lambda D. \dfrac{\dfrac{W}{C}}{\text{call-attendant name}} D \right\| \lambda D. \dfrac{C}{D \text{ octane}^\uparrow \text{ receipt?}} \right\| \text{credit-card} \quad \text{1234 93 call Joseph yes}$$

The current dialog is now atomic. If it does not match the first response, then staging fails. If there is a match, then the first response is removed, and the current dialog becomes the result of applying the top dialog constructor to the empty DSL expression. Again, simplification rules [EMPTY-1], [EMPTY-2], [EMPTY-3], and [EMPTY-4] are used, when possible (as shown below), where they are used to remove ∼ from underneath the *c* mnemonic.

$$\lambda D. \sim \left\| \lambda D. \dfrac{\dfrac{W}{C}}{\text{call-attendant name}} D \right\| \left( \lambda D. \dfrac{C}{D \text{ octane}^\uparrow \text{ receipt?}} (\sim) \right) \quad \text{93 call Joseph yes}$$

$$\lambda D. \sim \left\| \lambda D. \dfrac{\dfrac{W}{C}}{\text{call-attendant name}} D \right\| \dfrac{C}{\text{octane}^\uparrow \text{ receipt?}} \quad \text{93 call Joseph yes}$$

Once again, perform extraction on the current dialog.

$$\lambda D. \sim \left\| \lambda D. \dfrac{\dfrac{W}{C}}{\text{call-attendant name}} D \right\| \lambda D. \dfrac{C}{D \text{ receipt?}} \right\| \text{octane}^\uparrow \quad \text{93 call Joseph yes}$$

The current dialog is once again atomic. The $\uparrow$ operator only changes the number of dialog constructors to pop off the stack and apply, which is equal to one more than the number of arrows. With no arrows, only one dialog constructor is used, but in this case two are used.

$$\lambda D. \sim \left\| \left( \lambda D. \dfrac{\dfrac{W}{C}}{\text{call-attendant name}} D \left( \lambda D. \dfrac{C}{D \text{ receipt?}} (\sim) \right) \right) \right. \quad \text{call Joseph yes}$$

$$\lambda D. \sim \left\| \left( \lambda D. \dfrac{\dfrac{W}{C}}{\text{call-attendant name}} D \text{ (receipt?)} \right) \right. \quad \text{call Joseph yes}$$

$$\lambda D. \sim \left\| \dfrac{\dfrac{W}{C}}{\text{call-attendant name}} \text{receipt?} \right. \quad \text{call Joseph yes}$$

The current dialog once again uses the *w* mnemonic. This allows the other *c* sub-dialog to be selected for staging.

$$\lambda D. \sim \left\| \lambda D. \dfrac{W}{D \text{ receipt?}} \right\| \dfrac{C}{\text{call-attendant name}} \quad \text{call Joseph yes}$$

**The Formal Semantics and Implementation of a DSL for Mixed-Initiative Dialogs**

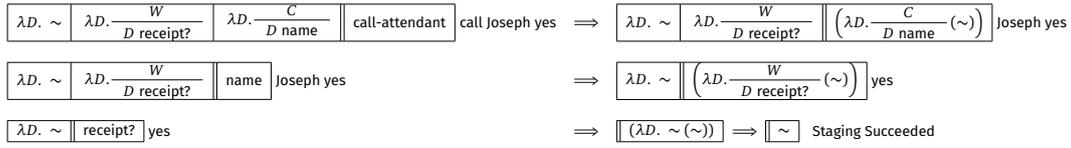

If both the stack and input string are empty, then staging succeeds.

## 3.4 Formal Semantics of DSL Expressions

We present a unified, single-step semantics of all concept mnemonics.

### 3.4.1 Definition of Formal Semantics of DSL Expressions

We use $\rightsquigarrow^\star$ to denote the reduction relation for the arrow extension in addition to simplification relation for dialogs in DSL. (The intended relation is clear from context.) The relation $\rightsquigarrow^\star \subset ((\mathcal{D} \to \mathcal{D})^\star \times \mathcal{D} \times \mathcal{N}^\star)^2$ is defined over the set of reduction states. A reduction state is a triple $(\Lambda, d, N)$, where $\Lambda$ is a stack of dialog constructors, $d$ is a dialog (e.g., an arrow expression), and $N$ is a sequence of user responses, each of which can be a single response $x$ or a set of responses $\{x^\star\}$. Unlike the semantics of DSL expressions without the $w$ mnemonic and $\uparrow$ operator, staging of dialogs that use arrows is defined using a single relation instead of separate simplification and staging relations.

The reduction relation $\rightsquigarrow^\star$ is defined as the reflexive and transitive closure of the single-step reduction relation $\rightsquigarrow$, which is in turn defined by a set of reduction rules (Appendix 3.4.2). The full set of reduction rules, including those for the core constructs of the arrow extension (i.e., [ATOM], [ARROW], [C], [W], [UNION-L], and [UNION-R]) is presented below. A star superscript denotes zero or more (e.g., $x^\star$), a plus superscript denotes one or more (e.g., $x^+$), and a subtraction symbol denotes set difference (e.g., $x^\star - y^\star$). The double colon (::) denotes the cons operator for lists, and nil denotes the empty list.

There are one or more rules for each type of DSL expression including atomic dialogs and dialogs with arrows. Since the $c$, $spe'$, and $w$ mnemonics allow sub-dialogs, their corresponding reduction rules describe how their sub-dialogs can be extracted. They grow the dialog-constructor stack and do not consume any input. Conversely, the rules [EMPTY], [ATOM], and [I] describe how input can be consumed and how constructors can be popped and applied. The [ARROW] rule handles arrows by collapsing the top two functions of the constructor stack into one function using composition. Multiple arrows on a single dialog are handled with successive applications of [ARROW]. Thus, $d^{\uparrow\uparrow}$ should be treated as $(d^\uparrow)^\uparrow$.

Note that the $\rightsquigarrow$ relation does not describe a deterministic staging function; it is possible for a triple to reduce in multiple ways. Most notably, triples where $d$ uses the $W$ mnemonic can be reduced in as many ways as there are sub-dialogs since each sub-dialog can be extracted. Unions of DSL expressions also use non-determinism to indicate that the dialog on the left or right of the $\cup$ can be discarded. However, only one reduction can lead to the completion of staging.

The relationship between dialogs and input strings can now be formally expressed using the reduction semantics. We say that an input string $N$ is a *member of* dialog $d$ (denoted $N \in d$) if there exists a sequence of reduction steps that results in the empty dialog and leaves the stack and input empty. This is expressed using the reflexive and transitive closure of the reduction relation, denoted $\leadsto^\star$. Formally,

$$N \in d \iff ((\text{const} \sim) :: \text{nil}, d, N) \leadsto^\star (\text{nil}, \sim, \text{nil})$$

It is also useful to express that a dialog $d$ can stage an input $N$, but results in a stack and dialog that can be used to stage further input. We say that $N$ is a *prefix of* dialog $d$ [denoted $\text{prefix}(N, d)$] if there exists reduction steps that consume the input regardless of the final dialog and stack. Formally,

$$\text{prefix}(N, d) \iff ((\text{const} \sim) :: \text{nil}, d, N) \leadsto^\star (\Lambda, d', \text{nil}) \text{ for some } \Lambda, d'$$

Like the reduction relation for DSL expressions without the $w$ mnemonic and $\updownarrow$ operator, $\leadsto^\star$ is a partial order and, therefore, must be an antisymmetric relation. As explicitly formulated, the reduction rules prevent antisymmetry—the empty dialog can be infinitely extracted and re-inserted because $\lambda D. \frac{C}{D\,x} :: \text{nil}, \sim, N) \leadsto (\text{nil}, \frac{C}{\sim x}, N)$ according to the [EMPTY] rule and $(\text{nil}, \frac{C}{\sim x}, N) \leadsto (\lambda D.\frac{C}{D\,x} :: \text{nil}, \sim, N)$ according to the [C] rule. This is resolved by requiring the output of dialog constructors to be simplified, which prevents the abuse of the [EMPTY] rule as demonstrated. For convenience, this simplification is not explicitly notated. All the other rules reduce the size of either the current dialog or the input list. Since the size of the input list never increases, repeated reduction should always eventually result in an irreducible expression (i.e., terminal). This argument is not sufficient as a formal proof of antisymmetry. However, it does intuitively justify the antisymmetric property of the relation.

### 3.4.2 Full Set of Reduction Rules For DSL Expressions

The definitions of the single-step ($\leadsto$) and full ($\leadsto^\star$) reduction relations that define staging for DSL expressions containing arrows are given here.

$$
\begin{array}{lll}
[\text{EMPTY}] & (f :: \Lambda, \sim, N) & \leadsto \quad (\Lambda, (f \sim), N) \\[4pt]
[\text{ATOM}] & (f :: \Lambda, x, x :: N) & \leadsto \quad (\Lambda, (f \sim), N) \\[4pt]
[\text{ARROW}] & (f_1 :: f_2 :: \Lambda, d^\updownarrow, N) & \leadsto \quad ((f_2 \circ f_1) :: \Lambda, d, N) \\[4pt]
[\text{UNION-L}] & (\Lambda, d_1 \cup d_2, N) & \leadsto \quad (\Lambda, d_1, N) \\[4pt]
[\text{UNION-R}] & (\Lambda, d_1 \cup d_2, N) & \leadsto \quad (\Lambda, d_2, N) \\[4pt]
[\text{C}] & \left(\Lambda, \dfrac{C}{d_1\, d_2^\star}, N\right) & \leadsto \quad \left(\lambda D.\dfrac{C}{D\, d_2^\star} :: \Lambda, d_1, N\right) \\[10pt]
[\text{W}] & \left(\Lambda, \dfrac{W}{d_1^\star\, d_2\, d_3^\star}, N\right) & \leadsto \quad \left(\lambda D.\dfrac{W}{d_1^\star\, D\, d_3^\star} :: \Lambda, d_2, N\right) \\[10pt]
[\text{I}] & \left(f :: \Lambda, \dfrac{I}{x^\star}, \{x^\star\} :: N\right) & \leadsto \quad (\Lambda, (f \sim), N) \\[10pt]
[\text{SPE}] & \left(\Lambda, \dfrac{SPE}{x_1^\star\, x_2\, x_3^\star}, x_2 :: N\right) & \leadsto \quad \left(\Lambda, \dfrac{I}{x_1^\star\, x_3^\star}, N\right) \\[10pt]
[\text{SPE}^\star] & \left(\Lambda, \dfrac{SPE^\star}{x_1^\star\, x_2\, x_3^\star}, x_2 :: N\right) & \leadsto \quad \left(\Lambda, \dfrac{SPE^\star}{x_1^\star\, x_3^\star}, N\right) \\
\end{array}
$$

# The Formal Semantics and Implementation of a DSL for Mixed-Initiative Dialogs

$$[\text{SPE}^\star] \quad \left(\Lambda, \frac{SPE^\star}{x^+}, \{x^+\} :: N\right) \quad \rightsquigarrow \quad (\Lambda, \sim, N)$$

$$[\text{SPE'}] \quad \left(\Lambda, \frac{SPE'}{d_1^\star\, d_2\, d_3^\star}, N\right) \quad \rightsquigarrow \quad \left(\lambda D.\frac{SPE'}{d_1^\star\, D\, d_3^\star} :: \Lambda, d_2, N\right)$$

$$[\text{PFA}_1] \quad \left(\Lambda, \frac{PFA_1}{x\, x_1^\star}, x :: N\right) \quad \rightsquigarrow \quad \left(\Lambda, \frac{I}{x_1^\star}, N\right)$$

$$[\text{PFA}_1^\star] \quad \left(\Lambda, \frac{PFA_1^\star}{x\, x_1^\star}, x :: N\right) \quad \rightsquigarrow \quad \left(\Lambda, \frac{PFA_1^\star}{x_1^\star}, N\right)$$

$$[\text{PFA}_1^\star] \quad \left(\Lambda, \frac{PFA_1^\star}{x^+}, \{x^+\} :: N\right) \quad \rightsquigarrow \quad (\Lambda, \sim, N)$$

$$[\text{PE}] \quad \left(\Lambda, \frac{PE}{x^\star}, \{x_s^\star\} :: N\right) \quad \rightsquigarrow \quad \left(\Lambda, \frac{I}{x^\star - x_s^\star}, N\right)$$

$$[\text{PE}^\star] \quad \left(\Lambda, \frac{PE^\star}{x^\star}, \{x_s^\star\} :: N\right) \quad \rightsquigarrow \quad \left(\Lambda, \frac{PE^\star}{x^\star - x_s^\star}, N\right)$$

$$[\text{LIFT}] \quad S_1 \quad \rightsquigarrow^\star \quad S_2 \quad \text{if } S_1 \rightsquigarrow S_2$$

$$[\text{REFLEXIVE}] \quad S \quad \rightsquigarrow^\star \quad S$$

$$[\text{TRANSITIVE}] \quad S_1 \quad \rightsquigarrow^\star \quad S_3 \quad \text{if } S_1 \rightsquigarrow S_2 \text{ and } S_2 \rightsquigarrow^\star S_3$$

## 3.5 Functional Programming Proof of Concept

To operationalize these formal semantics, we implemented the simplification and staging rules in Haskell.

### 3.5.1 Implementation of the Simplification and Staging Semantics

Haskell is ideally suited to operationalizing these formal semantics because pattern matching—the core of the rewrite rules—is an essential concept in Haskell. The source code for the implementation is available as a Git repository on GitHub[13] and Zenodo [40].

An architectural diagram of a dialog system implemented using DSL as the specification language is depicted in Figure 3 (bottom right). A code snippet of the staging implementation for DSL is in Listing 1. The grammar of DSL expressions is captured by the Dialog data type and user responses are represented as lists of strings. The reduction relations $\rightsquigarrow$ and $\rightsquigarrow^*$ and the staging function are represented by the functions simplify1, simplify, and stage, respectively. The simplify1 function applies a simplification rule, if possible, and simplify repeatedly simplifies the given DSL expression until canonical form is reached. The stage function applies the appropriate staging rule.

The nondeterminism of the single-step, simplification relation $\rightsquigarrow$ is not captured in simplify1. In the case where an expression can be simplified in one of multiple ways, the rule applied by simplify1 is arbitrary. However, due to confluence of $\rightsquigarrow^*$, this choice has no consequence on the canonical form of DSL expressions.

---

[13] https://github.com/saverioperugini/dsl (accessed 2025-01-21).

**Listing 1** Implementation of the DSL semantics.

```haskell
import Control.Applicative ((<|>))

data Dialog = Empty | Atom String | I [String] |
              C [Dialog] | SPEstar [String] |
              PEstar [String] |
              Union Dialog Dialog | ...

simplify1 :: Dialog -> Maybe Dialog
simplify1 d = empty1 d <|> empty2 d <|> empty3 d
         <|> empty4 d <|> atom1 d <|> atom2 d
         <|> flatten d

flatten :: Dialog -> Maybe Dialog
flatten (C ds) = C <$> flatten' ds
   where flatten' [] = Nothing
         flatten' ((C cs):ds) = Just (cs ++ ds)
         flatten' (d:ds) = (d:) <$> flatten' ds
flatten _ = Nothing

stage :: [String] -> Dialog -> Maybe Dialog
stage [y] (Atom x)
   | x == y = Just Empty
   | otherwise = Nothing
stage x (C (d:ds)) = (\d' -> C (d':ds)) <$> stage x d
   ...
```

**Listing 2** Implementation of arrow-reduction semantics.

```haskell
data Dialog = Empty | Atom String | I [String] |
              C [Dialog] | Up Dialog | ...

data Response = One String | Tup [String]
data RS =
   RS [Dialog -> Dialog] Dialog [Response]

reduce :: [RS] -> [RS]
reduce rs = case rs >>= reduce1 of
   [] -> rs
   rs' -> reduce rs'

reduce1 :: RS -> [RS]
-- [empty]
reduce1 (RS (f:lam) Empty inp) =
   [RS lam (f Empty) inp]
-- [atom]
reduce1 (RS (f:lam) (Atom x) ((One y):inp))
   = if x == y then [RS lam (f Empty) inp] else []
...
```

▪ **Listing 3**  Staging DSL expressions with arrows.

```
1  initDialog :: Dialog -> Maybe RS
2  initDialog d = Just (RS [const Empty] d [])
3
4  stage :: Response -> RS -> Maybe RS
5  stage inp (RS lam d _) =
6    case consumeOneInput [RS lam d [inp]] of
7      [] -> Nothing
8      [RS lam' d' []] -> Just (RS lam' d' [])
9      other -> Nothing
10   where consumeOneInput states =
11     states >>= (\state ->
12       case reduce1 state of
13       [] -> []
14       [RS lam' d' []] -> [RS lam' d' []]
15       other -> consumeOneInput other)
16
17 verifyComplete :: RS -> Maybe ()
18 verifyComplete (RS [] Empty []) = Just ()
19 verifyComplete _ = Nothing
20
21     initDialog (C [Atom "x", I ["y", "z"]])
22 >>= stage (One "a")
23 >>= stage (Tup ["y", "z"])
24 >>= verifyComplete
```

### 3.5.2 Implementation of the $w$ Mnemonic and Arrow Operator

The implementation of dialogs specified using the $w$ mnemonic extension and ↷ operator and their formal semantics is given in Listing 2. The RS datatype (for "reduction state") defines the set of $(\Lambda, d, N)$ triples acted on by the reduction rules. The single-step, reduction relation ⤳ is implemented by the function reduce1 that maps the given triple to the set of triples that may result after the application of one reduction rule. If a triple cannot be reduced further, reduce1 returns the empty list.

The implementation illustrates a consequential benefit of the triple-based reduction rules: the non-determinism of the reduction relation is handled entirely by the default monadic behavior of Haskell's list data type. For example, the expression [rs] »= reduce »= reduce computes all possible triples that can result from two reductions of rs without any explicit iteration.

The reduction rules are only defined for situations where the entire sequence of user responses is available. The reduction rules do not describe how a stager can alternate between reading a response and reducing the dialog expression, which is provided by the stage function in Listing 3. The function attempts to reduce the state until the response given as an argument is consumed. The resulting stack and dialog are returned on success. The stage function and two helper functions initDialog and verifyComplete abstract away the concept of the reduction state to enable ease of response-by-response staging. The monadic behavior of Maybe handles failure and provides a convenient notation for reducing a dialog by providing a response.

## 4 Theoretical Evaluation of the Expressiveness of DSL

In addition to the formal semantics of DSL presented in the previous section, the $w$ mnemonic and $\updownarrow$ operator represent the primary enhancement to DSL discussed in this article. In this section we are only concerned with evaluating the effect of the $w$ mnemonic and $\updownarrow$ operator—referred to here as the *arrow extension*—on how well DSL is capable of (losslessly) compressing an enumerated specification.[14] (Recall that an *enumerated (dialog) specification* is the complete set of episodes to be supported by a dialog system.) The arrow extension targets dialogs involving multiple concurrent sub-dialogs. Prior to the addition of the arrow extension, such dialogs could not be specified in DSL; to denote these dialogs we used a union of DSL expressions, where each expression contained only non-$w$ mnemonics, which involved an exponential blowup in the number of expressions required. Now such dialogs can specified in DSL:

$$\cfrac{\cfrac{\cfrac{W}{\text{credit-card-num}^{\updownarrow}\ \text{octane}^{\updownarrow}\ \text{receipt?}}\ \cfrac{C}{\text{call-attendant-for-help name}}}{} \equiv}{\cfrac{C}{\text{call-attendant name credit-card-num octane receipt?}} \cup \cfrac{C}{\text{credit-card-num call-attendant name octane receipt?}} \cup}$$
$$\cfrac{C}{\text{credit-card-num octane call-attendant name receipt?}} \cup \cfrac{C}{\text{credit-card-num octane receipt? call-attendant name}}$$

For purposes of comparison we refer to DSL without the $w$ mnemonic and $\updownarrow$ operator as *original DSL*, which represents the language in discussed in our prior work [7, 31, 33], and the language with those enhancements as *augmented DSL*. Given an enumerated specification $d$, we seek to measure the reduction in the minimum number of DSL expressions necessary to represent $d$ in augmented DSL ($t'$) vis-à-vis original DSL ($t$) (i.e., we seek to approximate $t - t'$).

Perugini [31] fosters a similar comparison between the number of episodes in an enumerated specification and the number of DSL expressions in original DSL that must be unioned to capture the episodes in that specification precisely (i.e., with no more or no less episodes). The comparison is made by using a dialog mining algorithm which tries to extract/discover the minimum number of DSL expressions that, when unioned, precisely capture the episodes in an (input) enumerated specification. The mining algorithm can alternatively be thought of as a compression algorithm that attempts to compress an enumerated specification into a DSL expression that compresses that specification to the highest degree. Out of the total 8,191 enumerated specifications constructed from three solicitations, 7,658 of them (93 %) were reduced in size after mining (i.e., 93 % of the specifications were compressible by codifying them in original DSL) [31], where the size of an enumerated specification is the number episodes in

---

[14] While a variety of methods exist for evaluating dialog systems [11, 25], we are evaluating a dialog-authoring language and not the resulting dialog systems built with augmented DSL.

**The Formal Semantics and Implementation of a DSL for Mixed-Initiative Dialogs**

it and the size of a union of a set of DSL expressions is the number of expressions unioned.

## 4.1 Methodology

In the absence of such a mining algorithm from enumerated specifications to augmented DSL expressions,[15] we only consider specifications for which the minimum number of augmented-DSL expressions is known *a priori* (i.e., without mining). One way of producing a set of such enumerated specifications is by first generating a synthetic set of augmented DSL expressions, and then decompressing each into the enumerated specification it captures (i.e., the inverse of dialog mining). Then, the mining algorithm can be used to generate the corresponding set of expressions in original DSL from them that capture them [31].

Thus, we first generated a set of 1,000 random expressions in augmented DSL, called $L_3$, as synthetic test data. Each expression has five solicitations and uses $w$ as the root mnemonic.[16] (Use of each of the other mnemonics as the root mnemonic only reduces the proportion of the generated DSL expressions that contain an arrow.) Also, $L_3$ does not include expressions specified with the $I$, $PE^*$, or other mnemonics that specify utterances that each respond to multiple solicitations since the arrow extension does not compress enumerated specifications captured by such expressions.

Next, the set of expressions in $L_3$ were decompressed into the corresponding set of enumerated specifications, called $L_1$. The decompression was conducting using the `genEpisodes` function (see Git repository), which stages the DSL expression with utterances of a single response, one at a time. Inputs that cause failure are discarded, while inputs that successfully advance staging by one step are retained, and the new reduction state is recursively staged with all inputs. The `getReds` function performs the single-staging step, and the `genEpisodes` function iterates `getReds` until a fixpoint is reached.

Finally, the set of enumerated specifications $L_1$ was compressed using the mining algorithm into the corresponding set of expressions $L_2$ in original DSL. Computing the average number of DSL expressions that must be unioned in each DSL expression in each of the sets $L_2$ and $L_3$ (i.e., $verb(L) = \forall e \in L, \frac{1}{|L|} \sum size(e)$) provides a general feel for the verbosity in original and augmented DSL, respectively. We seek to measure an increase in the ratio between the verbosity of original and augmented DSL—a value we call the *compression factor* of the arrow extension. The compression factor from original to augmented DSL ($\frac{verb(L_2)}{verb(L_3)}$) as well as the compression factor from the set of enumerated specifications to original DSL ($\frac{verb(L_1)}{verb(L_2)}$) and augmented DSL ($\frac{verb(L_1)}{verb(L_3)}$) were used to evaluate the impact of the arrow extension.

---

[15] Designing a mining algorithm from enumerated specifications to augmented DSL expressions is a complex algorithmic problem in and of itself and, thus, beyond the scope of this work.

[16] The degree of compression resulting from use of the other mnemonics is measured in [31].

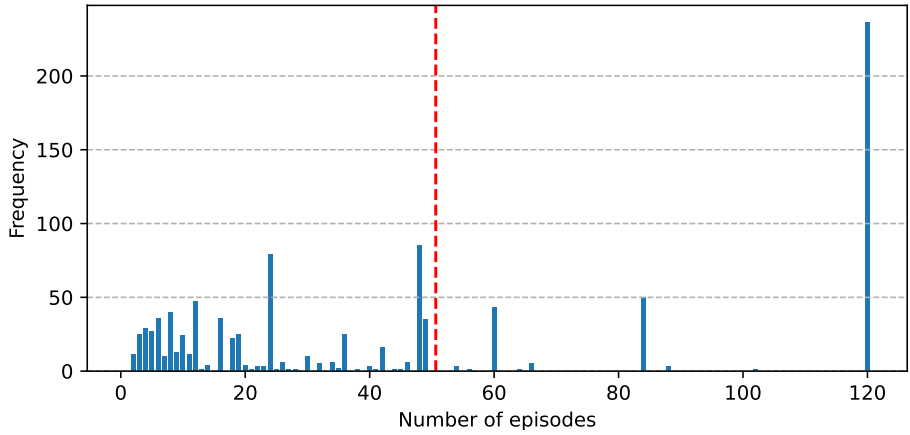

**Figure 6** Distribution of the sizes of the enumerated specifications in $L_1$. The verbosity of $L_1$ is shown by the red line ($verb(L_1) = 51$).

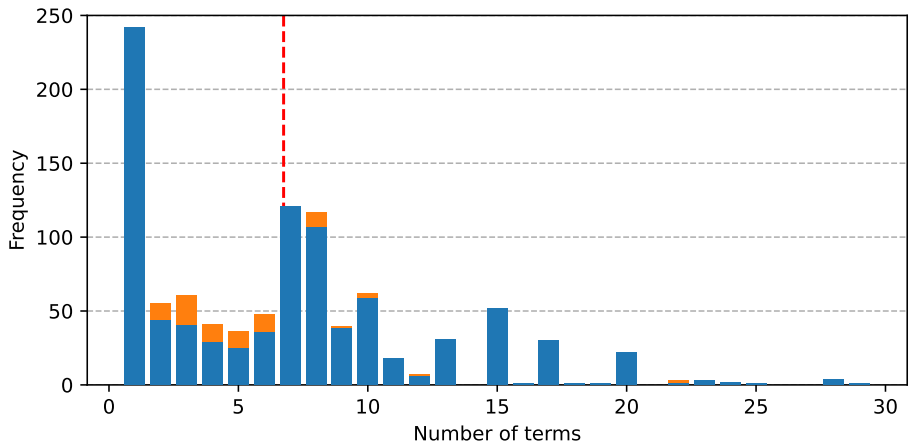

**Figure 7** Distribution of the sizes of the original DSL expressions in $L_2$. The verbosity of $L_2$ is shown by the red line ($verb(L_2) = 6.7$).

## 4.2 Results

Figures 6 and 7 contain histograms of the cardinality of the enumerated specifications in $L_3$ and the number of DSL expressions unioned in $L_2$, respectively. The verbosity (i.e., the average number of episodes in a generated dialog) of $L_1$ is 51. The verbosity of $L_2$ (after applying the mining algorithm) is 6.7 original DSL expressions, meaning that original DSL has a compression factor of 7.6 ($= \frac{51}{6.7}$). The verbosity of $L_3$ (after introducing the arrow extension) is 1.0 due to the method by which the dialogs were generated. That is, each of the 1,000 randomly generated dialogs were compressed to a single augmented-DSL expression. Therefore, the arrow extension offers an additional compression factor of 6.7 ($= \frac{6.7}{1.0}$) bringing the total compression factor from 7.6 to

51. These results indicate that the use of the arrow extension produces a significant decrease in the verbosity of the synthetic dataset.

We also computed the number of expressions that became compressible. The mining algorithm was able to compress 92 % of the enumerated specifications (represented by the blue bars in Figure 7). With the introduction of the arrow extension, this percentage increased to 100 %. In addition, the expression that underwent the greatest compression from original to augmented DSL was compressed from twenty-nine DSL expressions (unioned) to one expression.

### 4.3 Discussion

Using generation parameters, we sought to produce a synthetic test dataset of dialogs with uniform or binomial coverage of the set of DSL expressions. We fixed the probability that an atomic dialog is annotated with an arrow at 50 %, and parent dialogs are biased toward having an even distribution of atomic solicitations among its sub-expressions. Thus, the generated nature of the synthetic dataset implies that it is biased by these settings. Moreover, a general limitation of synthetic data is the uncertainty that it is representative of dialogs with concurrent sub-dialogs that would arise in practice.

Another limitation is that the mining algorithm, while sound,[17] is incomplete in that it does not always find the minimum number of DSL expressions that, when unioned, specify the (input) enumerated specification [31]. Thus, the compression factor is inflated since the output of the miner does not precisely reflect the expressive power of original DSL. Nonetheless, this theoretical evaluation provides informal evidence that the arrow extension is able to compress DSL expressions that were not compressible in [31].

All of the scripts and source code to run the experiments presented in the paper are available on Zenodo at https://doi.org/10.5281/zenodo.14717124 [40].

## 5 Related Research: Comparisons to Other Dialog Modeling Languages

The task-based approach to dialog design focuses on modeling a set of tasks to be supported by the system using a dialog-authoring language [5], and determining how to provide the user with desired interaction flexibility for completing those tasks, which can involve identifying the permissible (and impermissible) user-system sequences for completing the tasks. In this section, we draw comparisons between DSL and other formalisms that can be leveraged for representing sequences of user-system interaction in a dialog. We also make comparisons between DSL and two similar authoring languages: VoiceXML and State Chart XML.

---

[17] It will never output a DSL expression that does not capture the enumerated specification.

### 5.1 Theoretical Formalisms as Abstract Dialog Models

Transition networks (e.g., finite-state automata), context-free grammars, and event models—discussed in this subsection—have been used as abstract models to describe the structure of the (permissible) interaction in a human-computer dialog [15]. Green describes descriptive power as a measure of the expressiveness of a dialog model [15]. A model with more descriptive power can express a greater number and variety of dialog specifications. The descriptive power of a dialog-authoring language (e.g., DSL) can be evaluated by comparing the language to abstract models.

**Transition Networks and Context-Free Grammars** A transition network can be used as a general purpose dialog model because of its versatility and simple and intuitive visual representation (Figure 8, top right) [15]. Dialog specifications described by transition networks are suited for highly structured interactions where user initiative is limited and the set of valid user utterances is small [5, 27]. As the number of paths through the dialog and user responses increase, dialog specifications based on transition networks quickly grow large in size and increase in complexity.

While related, our model is not the same as a transition network or operators from regular expressions [20]. While currying and partial evaluation correspond independently to the *sequencing* and *shuffling* operators in (extended) regular expressions, the $w$ mnemonic, unlike shuffle, globally affects a dialog with nested sub-dialogs. While one can shuffle more than one string at once, the shuffling process remains linear in the sense that no string is subordinate to another. This contrasts with our model which imposes hierarchical dependencies between sub-dialogs which can then be temporarily violated with the $w$ mnemonic and ↥ operator.

Also, while the shuffle $ab \parallel cd$ denotes the language {$abcd, acbd, acdb, cabd, cadb, cdab$}, if we want to exclude the string cdab, we must consider a shuffle followed by a subtraction (i.e., $(ab \parallel cd) - \{cdab\}$). As we introduce additional restrictions on the episodes in a dialog specification, the (extended) regular expression capturing it becomes unwieldy.

Dialog systems typically accomplish goals through a divide-and-conquer technique: the dialog goal is divided into sub-goals, which are divided further into sub-sub-goals, and so on until all goals are expressed as a combination of atomic goals. Thus, since dialogs can contain arbitrarily nested sub-dialogs, transition networks are less effective as general discourse structures than other models [13]. Concepts from programming languages, such as currying, partial application, and others can help succinctly capture these nuanced requirements, particularly when dealing with compositions/nestings of sub-dialogs.

Context-free grammars (CFGs) can model richer dialogs and are particularly useful for describing sequences of user-system interaction that is hierarchical, or modular, in structure; can be described as a composition of sub-dialogs; or requires unbounded memory [13]. A transition network, in contrast, is monolithic and does not capture such properties. Transition networks, however, are often augmented with other features to provide more descriptive power. For instance, a *recursive transition network* (RTN) consists of multiple finite-state automata that can transition recursively among themselves. An RTN has the benefits of a stack-based approach akin to ours while

**The Formal Semantics and Implementation of a DSL for Mixed-Initiative Dialogs**

retaining the precise control offered by state machines. TuTalk, a manager for tutor dialogs, is implemented as a recursive transition network [22].

Abstract dialog models beyond the descriptive power of context-free grammars are modifications of less powerful models such as finite-state automata. For example, an *augmented transition network* is a recursive transition network (RTN) coupled with a set of registers and register functions [15]. The registers are updated when a state transition is taken with the primary state of the RTN.

While the properties of these formalisms are well established, and can be used to prove mathematical proprieties, interactions between the computer and user often need to be over-specified to support a flexible form of human-computer dialog. Similarly, CFGs might be appropriate if the utterances that participants in a dialog are going to say could be known, or predicted with high accuracy, in advance [13].

**Comparisons to DSL**  Concepts of programming languages like currying and partial evaluation are helpful metaphors in dialog design because they have analogs in human-human dialog and user interfaces (e.g., interruption). For instance, a system-initiated dialog like a multi-argument function (or an entire computer program) has a default, fixed flow of control. Currying and partial evaluation support the programmer in circumventing that default control flow in a function (or an entire computer program) in real-time. They do the same in a dialog; they support an unanticipated user-initiated interaction (e.g., an interruption). When we mix the two, we get a mixed-initiative style of interaction [30].

While other formalisms (e.g., extended regular expressions or CFGs) can be used to represent sequences of human-computer interactions in a dialog, the connection that a dialog has a default control flow (like a computer program) that can be circumvented by the user based on programming language theory that support unique types of circumventions is fundamentally different. A finite automata is fixed and cannot be altered at run-time. You could hard-wire every sequence into an finite automata or context-free grammar to make the dialog maximally flexible, but in some sense that is the equivalent to a union of factorially many $c$ mnemonics, which is what we are trying to avoid in our work.

**Event-Driven Models**  Green shows that event(-driven) models have the greatest descriptive power [15]. This is because an event model is embedded in a programming language, which gives the event model the descriptive power of a Turing machine. There are also a variety of well-established process calculi, including *Communicating Sequential Processes* [18] and the *Actor model* of concurrency [1, 17] that date back to the mid 1970s, that are also event-driven models.[18] Event-driven systems operate by concurrently executing a set of event handlers, each of which listens and responds to certain types of events. An event could be externally triggered by a user action or be internally generated by an event handler.

---

[18] There are also a variety of abstract models that involve parallel composition of automata (e.g., Petri nets, process algebras).

**Comparisons to DSL**    We explored a process-based, messing-passing model to describe the semantics of the *w* mnemonic. A process-based semantics aligns well with the intuitive behavior of arrows. Each mnemonic and atomic dialog in an expression corresponds to a process. The process for a parent dialog is responsible for managing the execution of its children, and each atomic process performs the work of making solicitations and receiving responses. This approach yields simple semantics: a dialog annotated with arrows, upon completion, passes control to a more distant ancestor rather than to its parent. However, using concurrent processes to implement staging was cumbersome because it is challenging to control the progression of the process-based staging. Rewinding a dialog to a previous state is particularly challenging vis-à-vis rewrite rules, where the dialog term can be simply saved after each reduction rule is applied. Moreover, this approach uses an error message to handle the non-determinism of staging dialogs involving the *w* mnemonic, which should be factored out of the formalization. The two dialog-authoring languages we expounded on next are event-driven models.

### 5.2  VoiceXML

VoiceXML is a w3c standard for describing interactive, voice-response (IVR) systems in XML. A VoiceXML document is a collection of forms. A *form* contains form items, which can be *fields,* where the user must respond to a prompt, or *blocks,* which specify system output without user input. Forms can transition control to other forms.

A form in VoiceXML is similar to the $PE^*$ mnemonic because both specify unrestricted slot-filling—the user can respond to a set of solicitations in any order and any combination. Likewise, a chain of forms, in which each form specifies another form as a successor, corresponds to the $c$ mnemonic because both specify that a set of dialogs be completed in a fixed order.

More complex dialogs, including those requiring the *w* mnemonic and ↱ operator, can be represented using the XML condition attribute (cond) for form items, which restricts the order in which the form items can be visited. For example, the condition ¬call-attendant ∨ name placed on the credit-card field means that credit-card can only be visited if call-attendant has not been visited or name has been visited. This condition is consistent with the absence of an ↱ operator after the call-attendant task in the sample DSL dialog given previously.

Figure 8 (bottom) presents a VoiceXML specification and the conditions for each task corresponding to the given DSL expression (left center). DSL offers multiple advantages in concision and readability over VoiceXML. The need to specify conditions for the completion of each task is completely obviated by the *w* mnemonic and ↱ operator. More importantly, DSL abstracts the control flow of the dialog away from task-specific properties such as the task prompt, method of collecting user input (e.g., DTMF vis-à-vis voice recognition), and error-handling procedures. In contrast, VoiceXML mixes control flow specification with task-specific logic. Finally, even when task-specific properties are omitted from a VoiceXML specification (as is the case in Figure 8—bottom), the notation of DSL is more concise than its VoiceXML analog.

# The Formal Semantics and Implementation of a DSL for Mixed-Initiative Dialogs

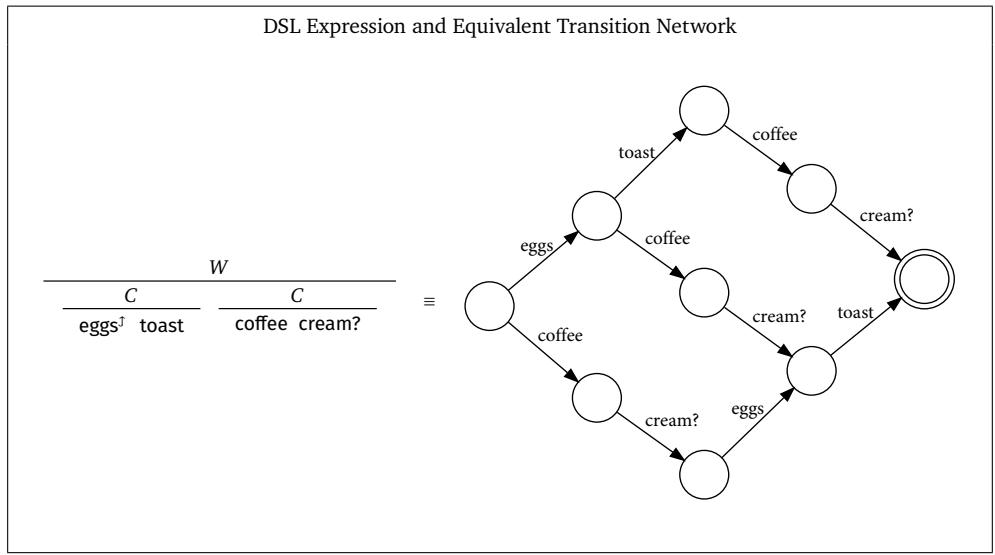

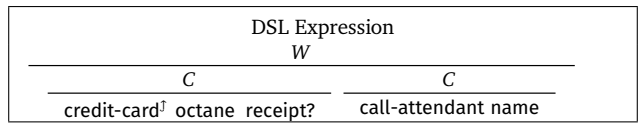

| Task | Condition |
|---|---|
| credit-card | $(\neg\text{credit-card} \vee \text{name})$ |
| octane | $(\neg\text{call-attendant} \vee \text{name}) \wedge \text{credit-card}$ |
| receipt? | $(\neg\text{credit-card} \vee \text{name}) \wedge \text{credit-card} \wedge \text{octane}$ |
| call-attendant | $(\neg\text{octane} \vee \text{receipt?})$ |
| name | $(\neg\text{octane} \vee \text{receipt?}) \wedge \text{credit-card}$ |

```vxml
<vxml>
  <form id="start">
    <field name="credit-card" cond="(!call-attendant || name)"></field>
    <field name="octane" cond="(!call-attendant || name) && credit-card"></field>
    <field name="receipt?" cond="(!call-attendant || name) && credit-card && octane"></field>
    <field name="call-attendant" cond="(!octane || receipt?)"></field>
    <field name="name" cond="(!octane || receipt?) && call-attendant"></field>
  </form>
</vxml>
```

■ **Figure 8** (top) A DSL expression with an equivalent transition network. (center) A DSL expression with (bottom) its VoiceXML equivalent.

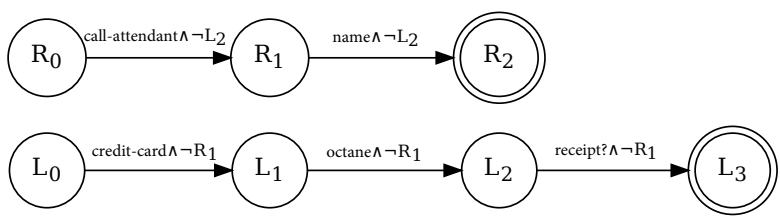

```
1  <scxml>
2    <parallel>
3      <state id="left">
4        <initial id="L0">
5          <transition event="credit-card" cond="!In('R1')" target="L1" />
6        </initial>
7        <state id="L1">
8          <transition event="octane" cond="!In('R1')" target="L2" />
9        </state>
10       <state id="L2">
11         <transition event="receipt?" cond="!In('R1')" target="L3" />
12       </state>
13       <state id="L3"></state>
14     </state>
15
16     <state id="right">
17       <initial id="R0">
18         <transition event="call-attendant" cond="!In('L2')" target="R1" />
19       </initial>
20       <state id="R1">
21         <transition event="name" cond="!In('L2')" target="R2" />
22       </state>
23       <state id="R2"></state>
24     </state>
25   </parallel>
26 </scxml>
```

**Figure 9** (top) A state-transition diagram and (bottom) the equivalent scxml specification corresponding to the dsl expression in Figure 8 (center).

A form is staged by the *form interpretation algorithm* (FIA), which repeatedly selects the first form item whose form item variable has not been filled and performs the actions associated with the form item (e.g., prompting for user input, reading user responses, and jumping to other sections of the dialog). The FIA is specified in prose and pseudocode, which increases the likelihood that different implementations of the algorithm deviate from each other and from the intended behavior. In contrast, the formal specification of the staging of DSL expressions, and the ease with which the formal specification is translated into a functional programming implementation (Sections 3.5 and 3.5.2), offer a higher level of security and correctness of any software implementations of the staging process.

Ramakrishnan, Capra, and Pérez-Quiñones [37] have demonstrated that the VoiceXML form interpretation engine is a partial evaluator in disguise. In comparison to our work, this means that while VoiceXML can support the episodes supported by a dialog specified with the *PE*\* mnemonic, it cannot as easily support those in the spectrum from *I* and *C* to *PE*\* as can DSL.

### 5.3 State Chart XML

State Chart XML (SCXML) is a W3C standard for expressing transition systems for use in voice applications [4]. SCXML is based on Statecharts which were introduced by Harel in 1987 as a visual formalism for describing state machines and their behavior in complex systems [16]. It extends the traditional transition network by introducing concepts like hierarchical, concurrent states and communication between them. SCXML aims to capture the similar concepts of structured and parallel states and event-based transitions between them. They differ in their syntactic representation of states (XML vis-à-vis graphical) and intended uses (web software applications vis-à-vis state-based modeling).

SCXML was designed to be compatible with VoiceXML, but can also be used to express transition systems for other purposes. Since some dialog systems use transition systems to manage control flow (e.g., Dialogflow[19] from Google), the improvements of DSL over SCXML apply to other dialog-authoring languages as well. Figure 8 (top) demonstrates how a dialog term in DSL can contain less boilerplate specification than an equivalent finite-state machine. The solicitations, which are modeled as transition labels, are repeated in the finite-state machine, whereas each solicitation is represented only once in the DSL expression.

States and transitions are represented using the <state> and <transition> elements, respectively, in SCXML. In addition, SCXML defines elements for richer transition systems. For instance, the <parallel> element specifies a set of sub-systems that are active concurrently. The <parallel> element is an analog of the *w* mnemonic.

Figure 9 (top) presents a state-transition diagram and (bottom) the equivalent specification in SCXML, both of which correspond to the DSL expression in Figure 8 (center). To represent a sample DSL dialog in SCXML, we convert the *w* mnemonic expression into a set of concurrent transition systems. Each sub-dialog under the *w* mnemonic becomes a transition system with a single active state that can transition independently of the other systems. A transition can only be taken after a certain task is completed and the current active states permit the transition. For instance, the condition for transitioning from $R_0$ to $R_1$ is call-attendant $\land \neg L_2$, which means the task call-attendant must be completed and the $L_2$ state must not be active. We then encode the set of parallel state-transition diagrams in SCXML syntax.

---

[19] https://cloud.google.com/dialogflow (accessed 2025-01-21).

DSL offers of the same benefits over SCXML as it offers over VoiceXML: the notation is more concise and control flow is separated from task-specific logic. The primary advantage of DSL over SCXML is that DSL expressions avoid explicitly capturing the superfluous details of the transition system (e.g., state names and transition conditions). State names are never given in any DSL expression, and transition conditions are determined by the locations of arrows. In summary, these two comparisons (to VoiceXML and State Chart XML) illustrate the advantages of DSL: the notation is more succint, eliminates boilerplate code/notation, and decouples the control flow from the dialog specification.

### 5.4 Other Programming Language-Based Approaches

Curried functions [32, Section 8.3], partial evaluation [21], and other programming language concepts have been used to specify dialogs, while term rewriting [3] has been applied to implement the semantics of those specifications. Use of concepts from programming languages in this manner represents a fundamentally different approach to modeling, managing, and implementing mixed-initiative dialogs. Program transformations, language concepts, and rewrite rules have also been utilized in peripherally related contexts to enable expressive forms of human-computer interaction. For instance, researchers have applied program slicing [38] and source-to-source rewrite rules [39] to restructure web interactions. Additionally, Quan, Huynh, Karger, and Miller [35] have investigated the use of currying and continuations to defer, save, and resume dialogs within application software. Queinnec [36] utilized first-class continuations to manage state in web dialogs, while Graunke, Findler, Krishnamurthi, and Felleisen [14] used them to automatically transform batch programs for interactive use on the web. A common theme in these efforts, in relation to our research, is the use of programming language concepts to design interactive computing systems. These concepts serve as the theoretical foundation for artistic implementation strategies.

## 6 Conclusion

In this work, we enriched DSL, a dialog-authoring language, with additional abstractions; formalized the semantics of DSL for dialog specification and implementation; operationalized DSL with a Haskell functional programming implementation; and evaluated the extension to DSL from practical (i.e., case study), conceptual (i.e., comparisons to similar systems), and theoretical perspectives.

# The Formal Semantics and Implementation of a DSL for Mixed-Initiative Dialogs

**Contributions** This research makes the following contributions to task-based, mixed-initiative dialog modeling and management: The the *w* mnemonic and ↕ operator enhance the expressiveness of DSL in modeling mixed-initiative dialogs and facilitates a concise representation of dialogs that interleave multiple sub-dialogs, similar to the behavior of coroutines; A formalization of the process of simplifying and staging a dialog specified in DSL [33]. In particular, the semantics were reformalized (i.e., simplification and staging rules were combined into one relation over the set of reduction states) and the semantics of the *w* mnemonic and ↕ operator were formalized using a reduction relation built from a set of reduction rules; A theoretical evaluation of the formal semantics, which provides informal evidence that the *w* mnemonic and ↕ operator improve the compression of DSL expressions previously reported [31]. Through random dialog generation and mining, we demonstrated that dialogs that could not be specified without a union of as many as twenty-nine DSL expressions could be compressed into a single DSL expression with the *w* mnemonic and ↕ operator; and A proof of concept, as an operationalization of the simplification and staging rules in Haskell, that the formal semantics are sufficient to implement a dialog system.

An interesting line of future work involves applying DSL for verifying properties of dialog specifications (e.g., the equivalence of two DSL expressions).

**Data Availability** All of the scripts and source code to run the experiments presented in the paper are available on Zenodo [40].

**Acknowledgements** We thank the three anonymous reviewers for their detailed and constructive feedback which helped improve the structure and contents of this article.

## A  Terms Used in This Article to Describe Dialog Specification Language

| Term | Definition or Description |
| --- | --- |
| **In General** | |
| dialog system | A conversational agent implemented in software. |
| dialog specification | A precise description of valid interaction sequences between a human and a computer system. |
| dialog-authoring language | A notational language for expressing dialog specifications. |
| **In Dialog Specification Language** | |
| solicitation | An atomic prompt for user input (e.g., size). |
| solicitation-set | The complete set of solicitations posed during a dialog. |
| utterance | A set of responses to a set of solicitations (e.g., {large, dark roast, with cream}). |
| episode | A complete sequence of solicitations (e.g., ≺size {blend, cream?} have-coupon?≻) |
| enumerated dialog specification | A set of episodes (e.g., {≺size blend cream?≻, ≺blend size cream?≻}). |
| mnemonic | Restricts the orders and groupings of responses to solicitations. |
| expression | An expression in the language of DSL (e.g., $\frac{C}{\text{size blend cream?}}$). |
| sub-expression | A DSL expression appearing in the denominator of a DSL expression. |
| atomic expression | A DSL expression that represents a dialog of one solicitation (e.g., size). |
| dialog term | An atomic expression or mnemonic expression. |
| canonical form | The DSL expression (of a dialog) in simplest form (e.g., $\frac{SPE'}{\text{size blend cream?}}$). |
| dialog staging | The function from a (DSL expression, utterance) pair to a DSL expression. |

## B  Formal Semantics of Dialog Specification Language Without Arrows

The semantics of DSL expressions without the $w$ mnemonic and $\updownarrow$ operator, including the definitions of the single-step ($\leadsto$) and full ($\leadsto^\star$) simplification relations and the staging function, are detailed here. The simplification and staging specifications use a meta-notation to indicate arbitrary dialogs of a particular structure. A $d$ matches any DSL expression, $x$ and $y$ match only questions, and $\mathcal{M}$ matches a concept mnemonic. The asterisk and plus superscripts are used to indicate a pattern that matches *at least zero* or *at least one* consecutive sub-elements, respectively (e.g., $d^\star$, $d^+$, $x^\star$, and $x^+$). Any of these patterns may also be sub-scripted to distinguish them from other patterns. Moreover, the staging rules use $(x^\star)$ and $(y^\star)$ to match arbitrary user responses and a subtraction symbol denotes set difference (e.g., $x^\star - y^\star$).

### B.1  Simplification Rewrite Rules

| | | | | |
|---|---|---|---|---|
| [EMPTY-1] | $\dfrac{\mathcal{M}}{}$ | $\leadsto$ | $\sim$ | $\forall \mathcal{M}$ |
| [EMPTY-2] | $\dfrac{\mathcal{M}}{d_1^\star \sim d_2^\star}$ | $\leadsto$ | $\dfrac{\mathcal{M}}{d_1^\star\ d_2^\star}$ | $\mathcal{M} \in \{C, SPE'\}$ |
| [EMPTY-3] | $\sim \cup\ d$ | $\leadsto$ | $d$ | |
| [EMPTY-4] | $d\ \cup \sim$ | $\leadsto$ | $d$ | |
| [ATOM-1] | $\dfrac{\mathcal{M}}{d}$ | $\leadsto$ | $d$ | $\mathcal{M} \in \{C, SPE'\}$ |
| [ATOM-2] | $\dfrac{\mathcal{M}}{x}$ | $\leadsto$ | $x$ | $\mathcal{M} \notin \{C, SPE'\}$ |
| [FLATTEN] | $\dfrac{C}{d_1^\star\ \dfrac{C}{d_2^\star}\ d_3^\star}$ | $\leadsto$ | $\dfrac{C}{d_1^\star\ d_2^\star\ d_3^\star}$ | |
| [LIFT] | $d$ | $\leadsto^\star$ | $d_r$ | if $d \leadsto d_r$ |
| [REFLEXIVE] | $d$ | $\leadsto^\star$ | $d$ | |
| [TRANSITIVE] | $d_1$ | $\leadsto^\star$ | $d_3$ | if $d_1 \leadsto^\star d_2$ and $d_2 \leadsto^\star d_3$ |
| [SUBSTRUCT-1] | $\dfrac{\mathcal{M}}{d_1^\star\ d\ d_3^\star}$ | $\leadsto^\star$ | $\dfrac{\mathcal{M}}{d_1^\star\ d_r\ d_3^\star}$ | if $d \leadsto^\star d_r$, for all $\mathcal{M}$ |
| [SUBSTRUCT-2] | $d \cup d_2$ | $\leadsto^\star$ | $d_r \cup d_2$ | if $d \leadsto^\star d_r$ |
| [SUBSTRUCT-3] | $d_1 \cup d$ | $\leadsto^\star$ | $d_1 \cup d_r$ | if $d \leadsto^\star d_r$ |

## B.2 Staging Rules

| | | | | |
|---|---|---|---|---|
| [ATOM] | $stage(x, x)$ | $=$ | $\sim$ | |
| [UNION] | $stage(d_1 \cup d_2, x)$ | $=$ | $stage(d_1, x) \cup stage(d_2, x)$ | if both exist |
| [UNION-L] | $stage(d_1 \cup d_2, x)$ | $=$ | $stage(d_1, x)$ | if $stage(d_1, x)$ exists |
| [UNION-R] | $stage(d_1 \cup d_2, x)$ | $=$ | $stage(d_2, x)$ | if $stage(d_2, x)$ exists |
| [I] | $stage\left(\dfrac{I}{x^\star}, (x^\star)\right)$ | $=$ | $\sim$ | |
| [C] | $stage\left(\dfrac{C}{d_1\ d_2^\star}, x\right)$ | $=$ | $\dfrac{C}{stage(d_1, x)\ d_2^\star}$ | if $stage(d_1, x)$ exists |
| [PFA$_1$] | $stage\left(\dfrac{PFA_1}{x_1\ x_2^\star}, x_1\right)$ | $=$ | $\dfrac{I}{x_2^\star}$ | |
| [PFA$_1^\star$-1] | $stage\left(\dfrac{PFA_1^\star}{x_1\ x_2^\star}, x_1\right)$ | $=$ | $\dfrac{PFA_1^\star}{x_2^\star}$ | |
| [PFA$_1^\star$-2] | $stage\left(\dfrac{PFA_1^\star}{x^\star}, (x^\star)\right)$ | $=$ | $\sim$ | |
| [PFA$_n$] | $stage\left(\dfrac{PFA_n}{x_1^\star\ x_2^\star}, (x_1^\star)\right)$ | $=$ | $\dfrac{I}{x_2^\star}$ | |
| [PFA$_n^\star$] | $stage\left(\dfrac{PFA_n^\star}{x_1^\star\ x_2^\star}, (x_1^\star)\right)$ | $=$ | $\dfrac{PFA_n^\star}{x_2^\star}$ | |
| [SPE] | $stage\left(\dfrac{SPE}{x_1^\star\ x_2\ x_3^\star}, x_2\right)$ | $=$ | $\dfrac{I}{x_1^\star\ x_3^\star}$ | |
| [SPE$^\star$-1] | $stage\left(\dfrac{SPE^\star}{x_1^\star\ x_2\ x_3^\star}, x_2\right)$ | $=$ | $\dfrac{SPE^\star}{x_1^\star\ x_3^\star}$ | |
| [SPE$^\star$-2] | $stage\left(\dfrac{SPE^\star}{x^\star}, \{x^\star\}\right)$ | $=$ | $\sim$ | |
| [SPE'] | $stage\left(\dfrac{SPE'}{d_1^\star\ d_2\ d_3^\star}, x\right)$ | $=$ | $\dfrac{C}{stage(d_2, x)\ \dfrac{SPE'}{d_1^\star\ d_3^\star}}$ | if $stage(d_2, x)$ exists |
| [PE] | $stage\left(\dfrac{PE}{x^\star}, \{y^\star\}\right)$ | $=$ | $\dfrac{I}{x^\star - y^\star}$ | if $y^\star \subset x^\star$ |
| [PE$^\star$] | $stage\left(\dfrac{PE^\star}{x^\star}, \{y^\star\}\right)$ | $=$ | $\dfrac{PE^\star}{x^\star - y^\star}$ | if $y^\star \subseteq x^\star$ |

## About the authors

**Zachary S. Rowland** is a software engineer at Tangram Flex, Inc. in Dayton, Ohio. Contact him at rowlandz09627@gmail.com.
https://orcid.org/0009-0009-3990-4617

**Saverio Perugini** is a Professor of Mathematics and Computer Science at Ave Maria University. Contact him at saverio.perugini@avemaria.edu.
https://orcid.org/0000-0002-1736-4009